\documentclass[aps,pre,twocolumn,bibnotes,groupedaddress]{revtex4-1}
\usepackage{amsmath,amsthm,amsfonts,amssymb,bm}
\usepackage{graphicx,subfigure,psfrag,epsfig,epstopdf}
\usepackage[dvipsnames]{xcolor}
\usepackage[]{natbib}

\usepackage[colorlinks=true,citecolor=blue]{hyperref}

\usepackage{epsfig}
\usepackage{epstopdf}

\numberwithin{equation}{section}

\newcommand{\vecv}[1]{\bm{{#1}}}
\newcommand{\tens}[1]{\bm{{#1}}}

\newcommand{\partials}[2]{\partial_{#2}{#1}}

\newcommand{\be}{\begin{equation}}
\newcommand{\ee}{\end{equation}}
\newcommand{\bea}{\begin{eqnarray}}
\newcommand{\eea}{\end{eqnarray}}

\newcommand{\gdotbar} { \bar{\dot{\gamma}}}

\newcommand{\edotbar} { \bar{\dot{\epsilon}}}
\newcommand{\ebar} { \bar{\epsilon}}
\newcommand{\edot} { \dot{\epsilon}}

\newcommand{\sigmae} { \sigma_{\rm E}}

\newcommand{\considere} {Consid\`ere }

\newcommand{\taur}{\tau_R}
\newcommand{\taud}{\tau_d}
\newcommand{\taus}{\tau_s}

\newcommand{\conf}{\tens{W}} 
\newcommand{\total}{\tens{T}}
\newcommand{\visc}{\tens{\Sigma}}

\begin{document}
\title{Necking after extensional filament stretching of complex fluids
  and soft solids}

\author{D. M. Hoyle} 
\email{d.m.hoyle@durham.ac.uk}
\homepage{http://community.dur.ac.uk/d.m.hoyle/ }
\affiliation{Department of Physics, University of Durham, Science
	Laboratories, South Road, Durham, DH1 3LE, United Kingdom}

\author{S. M. Fielding} 
\affiliation{Department of Physics, University of Durham, Science
  Laboratories, South Road, Durham, DH1 3LE, United Kingdom}

\begin{abstract}
  We perform linear stability analysis and nonlinear slender filament
  simulations of extensional necking in complex fluids and soft
  solids, during the stress relaxation process following an
  interrupted strain ramp.  We start by deriving analytical criteria
  for necking within a highly simplified and generalised scalar
  constitutive model. Within this, we find two different possible
  modes of necking: one associated with an upward curvature in the
  stress relaxation function on a log-linear plot, and another related
  to a carefully defined `elastic' derivative of the tensile force
  with respect to an imagined sudden strain increment. We showed these
  two criteria to agree fully with simulations of the Oldroyd B and
  Giesekus models of polymeric solutions, and with the Rolie-Poly
  model of more concentrated polymeric solutions and melts, without
  polymer chain stretch.  With chain stretch included, we find a
  slightly more complicated analytical criterion for necking during
  the stress relaxation, although with key ingredients that closely
  mirror counterpart ingredients of the simpler criteria obtained
  within the scalar model.  We show this criterion to agree fully with
  slender filament simulations of the Rolie-Poly model with chain
  stretch, and with the scenario discussed by the Copenhagen group in
  Refs.~\cite{Lyhne2009,Rasmussen2011}.  In particular, we see delayed
  necking after strain ramps with an accumulated strain exceeding
  $\ebar\approx 0.7$, for ramp rates exceeding the inverse chain
  stretch relaxation timescale. We discuss finally an analogy between
  this delayed necking following an interrupted extensional strain
  ramp and delayed shear banding following an interrupted shear strain
  ramp~\cite{Moorcroft2013}. This work provides the counterpart, for
  interrupted extensional strain ramps, to earlier papers giving
  criteria for necking in the protocols of constant imposed Hencky
  strain rate~\cite{Hoyle2016a} and of constant imposed tensile stress
  or constant imposed tensile force~\cite{Hoyle2016}.

\end{abstract}

\date{\today}
\maketitle

\section{Introduction}
\label{sec:Intro}

Extensional flows provide an important test of the constitutive
properties of complex fluids and soft solids such as
linear~\cite{Auhl2008a}, star~\cite{Huang2016b},
branched~\cite{Barroso2005a,Liu2013} and associative
polymers~\cite{Tripathi2006,Huang2016a}, wormlike micellar
surfactants~\cite{Bhardwaj2007}, bubble rafts~\cite{Arciniaga2011} and
colloidal suspensions~\cite{Smith2010}. Typically, they subject the
underlying material microstructure (polymer chains, wormlike micelles,
foam bubbles, etc.) to much more severe reorganisation than is
experienced in shear. As a result, many nonlinear flow phenomena
manifest themselves only in extension. An obvious example is the
strain hardening seen in extensional flows of polymeric
fluids~\cite{Rolon-Garrido2009,Liu2013a}, compared with thinning in
shear. Extensional flows therefore provide an important way of
characterising a material's underlying microstructure and molecular
architecture~\cite{Munstedt2005,Lentzakis2013}, and of discriminating
between alternative constitutive theories.  For reviews,
see~\cite{McKinley2002b,Malkin2014}.

Perhaps the most common extensional flow experiment consists of taking
an initially cylindrical filament (or planar sheet) and stretching it
out in length under conditions of constant imposed Hencky strain
rate~\cite{Rasmussen2005a,Burghelea2011a,Rolon-Garrido2014,Wagner2013}.
Other commonly used protocols are those of constant imposed
tensile stress~\cite{Alvarez2013,Wolff2011}, constant imposed tensile
force~\cite{Szabo2012,Wagner2012}, large amplitude oscillatory
extension~\cite{Rasmussen2007}, or a ramp of finite strain amplitude
that is then interrupted~\cite{Bhattacharjee2003, Wang2007, Wang2008,
  Wang2011, Rasmussen2011, Huang2012, Hawke2015, Huang2016}.  The
usual aim in any such experiment is to draw the sample out in as
uniform a way as possible, in order to characterise its homogeneous
flow response.  Almost ubiquitously observed, however, is the
phenomenon of
necking~\cite{Barroso2005a,Liu2013,Burghelea2011a,Arciniaga2011,Smith2010,Andrade2011,Malkin2014,Wang2007},
in which some region along the filament's length thins more quickly
than the rest, forming a neck. This often leads the filament to fail
altogether at the neck, aborting the experiment.

Necking has been studied theoretically in
Refs.~\cite{Considere1885,Hutchinson1977,Ide1977,Malkin1997a,Olagunju1999,Olagunju2011,Joshi2003,Joshi2004,Hassager1998,McKinley1999a,Rasmussen2013,Cromer2009,Eastgate2003,Hoyle2016a,Hoyle2016}.
Much early work was based on the \considere criterion for necking in
solids~\cite{Considere1885}, which predicts instability to necking in
any regime where the tensile force $F$ in the filament is a declining
function of the accumulated Hencky strain $\epsilon$.  However, in
failing to take account of the {\em rate} of extension, this criterion
is unable to address necking in viscoelastic materials, in which
the rate at which the strain is applied (compared with the material's
intrinsic rate of stress relaxation $1/\tau$) is an important
variable, alongside the total accumulated strain.

Motivated by this shortcoming, we recently provided criteria for
necking in viscoelastic fluids and soft solids, separately for the
protocols of constant imposed Hencky strain
rate~\cite{Fielding2011,Hoyle2015,Hoyle2016a}, constant imposed
tensile stress~\cite{Hoyle2016}, and constant imposed tensile
force~\cite{Hoyle2016}.  These criteria were derived analytically
within a simplified constitutive model of highly generalised form,
with the aim that they should be as fluid-universal as possible,
independent of the particular assumptions of any given constitutive
model. They were then confirmed against numerical simulations of
several of the most widely used constitutive models of dilute polymer
solutions, concentrated solutions and melts of entangled linear and
branched polymers, wormlike micellar surfactants, and soft glassy
materials.

In the present manuscript, we turn to another important filament
stretching protocol: that of an interrupted extensional ``strain
ramp''~\cite{Bhattacharjee2003, Wang2007, Wang2008, Wang2011,
  Rasmussen2011, Huang2012, Hawke2015, Huang2016}.  In this protocol,
an initially uniform cylindrical filament (or planar sheet) is subject
to the switch-on at some time $t=0$ of a constant Hencky strain rate
$\edotbar$. (The overbar signifies the nominal strain rate averaged
along the filament's length.  Should the flow become heterogeneous,
the actual strain rate will vary as a function of position locally
along the filament.) The filament length then increases as
$L(t)=L(0)\exp(\edotbar t)$.  At some time $t_0$ the strain is
switched off and the sample is held in its strained state with new
length $L(t_0)=L(0)\exp(\edotbar t_0)$ for all subsequent times. The
tensile stress that accumulated during the initial straining process
then progressively relaxes as a function of the time since the
straining stopped. Indeed, this test is widely used to investigate a
material's stress relaxation
properties~\cite{Bhattacharjee2003,Sentmanat2005,Nielsen2008a,Huang2012,Hawke2015,Huang2016}
and the associated damping function~\cite{Barroso2002,
  Rolon-Garrido2009}.

Observed in many cases during the process of stress relaxation after
the straining has stopped, however, is a delayed necking instability,
which often leads the filament to fail entirely
~\cite{Sentmanat2005,Nielsen2008a,Huang2012,Wang2007,Wang2008,Wang2011}.
Any such failure clearly hampers attempts to characterise a material's
stress relaxation properties: when the sample fails the stress falls
catastrophically and the measurement terminates.

This phenomenon of delayed necking after an interrupted extensional
strain ramp was studied by numerical simulation in the insightful
earlier work of Refs.~\cite{Lyhne2009,Rasmussen2011}. These studies
showed that the effect could be completely accounted for within
continuum models of polymer rheology, based on long-standing concepts
of tube dynamics~\cite{Doi1986}, provided these are augmented to
include the concept of chain stretch~\cite{Fang1999, Mead1998,
  McLeish1998, Schieber2003}. Important physical conclusions of the
work were that (i) during the initial straining process, the samples
could be extended to nominal Hencky strains $\ebar\approx 2$ without
necking, (ii) after the straining stops, samples that had been
strained up to $\ebar\approx 0.6$ would remain intact, avoiding any
necking instability, while (iii) any samples that had been strained
beyond $\ebar\approx 0.8$ would show delayed necking after the
straining stops, and finally that (iv) this delayed necking sets in on
a timescale much shorter than the material's terminal reptation time.

The aim of the present manuscript is to build on the simulation
results of Ref.~\cite{Lyhne2009,Rasmussen2011} by performing
analytical linear stability calculations complemented by nonlinear
simulations of necking in complex fluids and soft solids, for this
interrupted strain ramp protocol. In particular, we shall analytically
derive criteria for the onset of necking after the end of the
straining process. We do this first within a simplified constitutive
model written in a highly generalised form, with the aim that the
criteria we provide are as fluid-universal as possible, independent of
the assumptions of any particular constitutive model.  We then show
the criteria to be in excellent agreement with the behaviour of the
Oldroyd B and Giesekus models of polymer solutions, and the Rolie-Poly
model of concentrated solutions and melts of entangled linear
polymers, and wormlike micellar surfactant solutions, provided chain
stretch is ignored.

We then proceed to incorporate polymer chain stretch.  With this
included, we find a slightly more complicated analytical criterion
for the onset of necking. However the key ingredients of this updated
criterion closely mirror counterpart ingredients of the
simpler criteria derived within the generalised model (and checked
against Oldroyd B, Giesekus and non-stretch Rolie-Poly).  We show that
this new criterion agrees fully with our numerical simulations of the
Rolie-Poly model with chain stretch included, and with the conclusions
of Refs.~\cite{Lyhne2009,Rasmussen2011}: in
particular in predicting delayed necking after fast strain ramps with
a total accumulated strain exceeding $\ebar\approx 0.7$.  We also
perform a quantitative comparison of our simulations with the
experiments of Ref.~\cite{Wang2007}, demonstrating excellent
agreement. Finally, we elucidate a close analogy between this scenario
of delayed necking following an interrupted extensional strain ramp
and that of delayed shear banding following an interrupted shear
strain ramp, as first put forward by one of the present authors
together with Moorcroft in Ref.~\cite{Moorcroft2013}.

Throughout we focus on the case of a highly viscoelastic filament of
sufficiently large initial cross sectional area that surface tension
can be safely ignored in comparison with the bulk viscoelastic
stresses, at least in considering the initial stages of neck
formation. In this way, we ignore any filament breakup and beading
instabilities driven by surface
tension~\cite{Clasen2006,Tembely2012,Vadillo2012,Webster2008,McIlroy2014}.

The paper is structured as follows. In Sec.~\ref{sec:models} we
introduce the continuum models to be studied.  Sec.~\ref{sec:protocol}
defines the flow protocol and geometry.  In Sec.~\ref{sec:linear} we
perform a linear stability analysis to derive criteria for the initial
onset of necking, and show them to be in excellent agreement with
numerical simulations of several of the most commonly used
constitutive models of polymeric fluids.  Sec.~\ref{sec:nonlinear}
reports our simulations of nonlinear necking dynamics, once the
necking perturbations have grown to be no longer small. In particular,
we quantitatively compare our calculations with the experiments of
Ref.~\cite{Wang2007}, demonstrating excellent agreement.  In
Sec.~\ref{sec:banding} we discuss the analogy with shear banding after
an interrupted shear strain ramp.  Sec.~\ref{sec:conclusion} contains
our conclusions and outlook for future work.

While this manuscript is intended to be self contained in its own
right, it would best be read after our earlier manuscripts in
Refs.~\cite{Hoyle2016a,Hoyle2016}. In making this manuscript
self-contained, the discussion in some places (particularly the
earlier introductory sections) inevitably mirrors that of those
earlier papers~\cite{Hoyle2016a,Hoyle2016} to some degree.

\section{Rheological models}
\label{sec:models}

We assume the stress $\total(\tens{r},t)$ in a fluid element at
position $\tens{r}$ at time $t$ to comprise a Newtonian solvent
contribution with viscosity $\eta$, an isotropic contribution from a
pressure field $p(\tens{r},t)$, and a viscoelastic contribution
$\visc(\tens{r},t)$ arising from the internal fluid microstructure
(polymer chains, wormlike micelles, {\it etc.}), giving:
\be
\total = \visc + 2 \eta \tens{D} - p\tens{I}.
\label{eqn:total_stress_tensor}
\ee
In this expression, $\tens{D} = \frac{1}{2}(\tens{K} + \tens{K}^T)$ is
the symmetrised strain rate tensor, with $K_{\alpha\beta} =
\partial_{\alpha}v_{\beta}$, and $\tens{v}(\tens{r},t)$ is the
velocity field of the fluid flow.  We assume creeping flow in which
the condition of force balance gives:
\be
\vecv{\nabla}\cdot\,\total = 0.
\label{eqn:force_balance}
\ee
We also assume the flow to be incompressible, with the pressure field
$p(\tens{r},t)$ determined by the requirement that the velocity field
remains divergence free:
\be
\label{eqn:incomp}
\vecv{\nabla}\cdot\vecv{v} = 0.
\ee

The dynamics of the viscoelastic stress $\visc$ comprises loading
driven by the velocity gradients of any imposed flow, combined with
relaxation back towards an unstressed equilibrium state. For any given
fluid, these two physical processes are governed by a rheological
constitutive equation. In this work we shall consider several
different constitutive models, which we specify now.

The phenomenological Oldroyd B model describes the rheology of a
dilute polymer solution by representing each polymer molecule as a
dumbbell formed of two beads linked by a Hookean spring.  The
viscoelastic stress 
\be
\label{eqn:Hooke}
\visc = G\left( \conf - \mathbf{I} \right),
\ee
in which $G$ is a constant modulus and $\conf=\langle
\vecv{R}\vecv{R}\rangle$ is a conformation tensor formed from the
ensemble average of the outer dyad of the dumbbell end-to-end vector
$\vecv{R}$. This is assigned dynamics in flow as follows:
\be
\overset{\nabla}{\mathbf{W}} = -\frac{1}{\tau} \left( \mathbf{W} - \mathbf{I} \right),
\label{eqn:Maxwell}
\ee
with a relaxation time $\tau$. The upper convected derivative
\be
\label{eqn:UCD}
\overset{\nabla}{\mathbf{W}} = \frac{\partial \conf}{\partial t} + \mathbf{v}\cdot\nabla\conf - \conf\cdot\mathbf{K} - \mathbf{K}^T\cdot\conf.
\ee

The Giesekus model describes more concentrated polymer solutions by
incorporating an anisotropic drag characterised by a parameter
$\alpha$, with $0 \le \alpha\le 1$. This encodes the basic idea that
the relaxation dynamics of any dumbbell is altered when its
surrounding dumbbells are oriented~\cite{Larson1988}, giving the
modified dynamical equation
\be 
\overset{\nabla}{\conf} = - \frac{1}{\tau} \left( \conf
  - \mathbf{I} \right) - \frac{\alpha}{\tau} \left( \conf -
  \mathbf{I} \right)^2.
\ee
The Oldroyd B model is recovered for $\alpha=0$.

The Rolie-Poly model~\cite{Likhtman2003} describes more concentrated
solutions or melts of entangled linear polymers, or solutions of
wormlike micellar surfactants.  It is based on the tube theory of
polymer dynamics~\cite{Doi1986}, in which any given polymer chain (or
wormlike micelle) is assumed to be dynamically restricted by
entanglements with its surrounding chains (or micelles).  Over time
the chain refreshes its configuration by a process of `reptation',
i.e., curvilinear diffusion back and forth along the tube contour, on
a timescale $\taud$.  An applied flow also induces stretch of any
chain along its tube. This relaxes on a timescale $\taus$, providing
an additional mechanism for relaxing entanglement points, known as
`convective constraint
release'~\cite{Marrucci1996,Ianniruberto2014,Ianniruberto2014a}. With
all these processes accounted for, the conformation tensor
$\conf=\langle \vecv{R}\vecv{R}\rangle$, in which $\vecv{R}$ is the
end-to-end vector of a polymer chain, has dynamics:
\begin{widetext}
\be
\overset{\nabla}{\conf} = - \frac{1}{\tau_d} \left( \conf - \mathbf{I} \right) 
  - \frac{2}{\taus}\left( 1 - \sqrt{\frac{3}{T}}\right)
\left[ \conf  + \beta\left(\frac{T}{3}\right)^{\delta}( \conf - \mathbf{I} ) \right].
\label{eqn:sRP}
\ee
\end{widetext}
In this equation, $T=\sum_i W_{ii}$ is the trace of the conformation
tensor. The parameter $\beta$ sets the degree of convective constraint
release, with $0\le\beta\le 1$.

The timescales of reptation and chain stretch relaxation are assumed
to be in the ratio
\be
\frac{\taud}{\taus}=3Z_{ent},
\label{eqn::3Z}
\ee
where $Z_{ent}$ is the number of entanglements per chain.  For highly
entangled chains, $Z_{ent}\gg 1$, reptation occurs much more slowly than the
relaxation of chain stretch: $\taud\gg\taus$. In this case, for flow
rates much less than $1/\taus$, we can use the simpler, non-stretching
form of the model:
\be
  \overset{\nabla}{\conf} 
= -\frac{1}{\taud}(\conf-\mathbf{I})-\frac{2}{3}\mathbf{K}:\conf\,\left[\conf+\beta(\conf-\mathbf{I})\right],
\ee
in which the limit $\taus\to 0$ has been taken upfront.

In both the Giesekus and Rolie-Poly models, the viscoelastic stress is
specified in terms of the conformation tensor by Eqn.~\ref{eqn:Hooke},
as in the Oldroyd B model. Throughout we use units in which the
modulus $G=1$, the relaxation time $\tau=1$ (Oldroyd B and Giesekus)
or $\taud=1$ (Rolie-Poly), and the initial filament length $L(0)=1$.

\section{Flow protocol and geometry}
\label{sec:protocol}

Before any flow commences, the filament is assumed to be prepared in
the shape of an undeformed uniform cylinder of length $L(0)=1$ in the
$z$ direction, with a cross sectional area $A(0)$ in the $xy$ plane.
All viscoelastic stresses are taken to be well relaxed, with
$\conf(0)=\tens{I}$.

At some time $t=0$, the filament is subject to the switch-on of a
Hencky strain rate $\edotbar$ that stretches the filament out along
the $z$ axis, with a flow field
\be
\bar{\tens{K}}=\edotbar\begin{pmatrix} -\frac{1}{2} & 0 & 0 \\
                           0 & -\frac{1}{2} & 0 \\
                           0 & 0 & 1 \\
	\end{pmatrix}.
\ee
(As noted above, the overbars signify that $\edotbar$ is the nominal
Hencky strain rate averaged along the length of the filament.  Once
any necking arises, the strain rate will vary locally as a function of
position $z$ along the filament's length.)  This strain rate
$\edotbar$ is held constant during a time interval $0<t<t_0$, with the
filament progressively drawing out in length according to
$L(t)=L(0)\exp(\edotbar t)$. Its cross sectional area thins
accordingly, to satisfy incompressibility.  During this straining
process, viscoelastic tensile stresses develop as a function of the
time $t$ since the inception of the flow.

After a time $t=t_0$, once a strain $\ebar_0=\edotbar t_0$ has
accumulated and the sample has attained a new length
$L(t_0)=L(0)\exp(\ebar_0)$, the strain rate is set back to zero and
the filament is held in this strained state with length $L(t_0)$ for
all times $t>t_0$, during which the viscoelastic stresses
progressively relax back to zero.

We shall call this procedure, comprising both the initial straining
and the stress relaxation that follows it, a `strain ramp protocol'.
Any such ramp is fully specified by the two parameters
$\ebar_0,\edotbar$, which respectively denote the total strain applied
$\ebar_0$ and the rate $\edotbar$ at which it is applied.  Clearly,
during the initial stretching process this strain ramp protocol
coincides with the simpler protocol in which a constant Hencky strain
rate $\edotbar=\rm{const.}$ is applied indefinitely, for all times
$t>0$ after the initial switch-on. In earlier papers, we considered
the dynamics of necking under those
conditions~\cite{Fielding2011,Hoyle2016a}.  The key difference in the
strain ramp protocol considered here is that the straining only
persists for a finite time $0<t<t_0$.

A key aim in what follows will be to determine whether, for any given
fluid and strain ramp, necking will arise at any stage during the
protocol. We further aim to determine whether any such necking will
occur primarily during the initial stretching, or primarily during the
stress relaxation that follows it (or roughly equally during both).
Clearly, any case in which necking mainly occurs during the initial
stretching process is already covered by our earlier results for the
protocol of a constant Hencky strain rate applied indefinitely after
the initial switch-ons~\cite{Fielding2011,Hoyle2016a}. Therefore, our
particular interest here will be in cases where necking is actually
suppressed during the initial stretching process itself, then occurs
with delayed onset during the subsequent stress relaxation.

Throughout we use a slender filament approximation \cite{Forest1990,
Olagunju1999, Denn1975}, which assumes the characteristic wavelengths
of any necking variations in cross sectional area that develop along
the filament's length to be large compared with the filament's radius.
This in turn allows the flow variables to be averaged over the
filament's cross section at any location $z$ along its length.  This
approximation has been carefully tested against full axisymmetric
simulations and shown to remain excellent even well into the nonlinear
regime, where the amplitude of perturbations becomes large and the
shear contribution might be expected to be large \cite{Vadillo2012}.
The dynamical variables to be considered are then the cross sectional
area $A(z,t)$, the area-averaged fluid velocity in the $z$ direction
$V(z,t)$, the extension rate $\edot(z,t)=\partial_z V$, and any
viscoelastic variables, as governed by the relevant rheological
constitutive equations set out in Sec.~\ref{sec:models}.  The
$z$-average of the local extension rate $\edot(z,t)$ is the nominal
Hencky strain rate $\edotbar$ defined above.

Within this slender filament approximation, the mass balance condition
(\ref{eqn:incomp}) becomes
\be
\partials{A}{t}(z,t) + V\partials{A}{z} = -\dot\varepsilon A. 
\label{eqn:1Dmass} 
\ee
The force balance condition (\ref{eqn:force_balance}) becomes
\be
0 = \partials{F}{z},
 \label{eqn:1Dmom} 
\ee
in which the tensile force
\be
\label{eqn:tforce}
F(t)=A(z,t)\sigmae(z,t),
\ee
and the total tensile stress 
\be
\label{eqn:tstress}
\sigmae = G\left(W_{zz} - W_{xx} \right) + 3\eta\dot\varepsilon.  \ee
The Lagrangian derivative of any constitutive model (first two terms
on the right hand side of Eqn.~\ref{eqn:UCD}) is written at this level
of slender filament as:
\be
\frac{D}{Dt}=\frac{\partial}{\partial t} + V\frac{\partial}{\partial z}.
\ee

In any given fluid, the viscoelastic stress $W_{zz}-W_{xx}$ in
Eqn.~\ref{eqn:tstress} (recall the modulus $G=1$ in our units) is
determined by the time-dependent components $W_{ij}(z,t)$ of the
appropriate tensorial constitutive equation of Sec.~\ref{sec:models}.
To allow analytical progress in our linear stability calculation of
Sec.~\ref{sec:linear} below, however, we shall initially consider a
simplified scalar constitutive model that denotes $W_{zz}-W_{xx}$ as a
single variable $Z$, for which it then postulates highly generalised
constitutive dynamics:
\be
\label{eqn:toy}
\frac{DZ}{Dt}=\edot f(Z)-\frac{1}{\tau}g(Z).
\ee
This has separate loading and relaxation terms characterised by the
functions $f$ and $g$ respectively.  In the first part of our analysis
below, we shall intentionally refrain from specifying any particular
functional forms for $f$ and $g$, in order that the criteria for
necking that we shall derive are as fluid universal as possible,
independent of the particular assumptions of any given constitutive
model. For notational simplicity in this scalar model, we also
renormalise the solvent stress $\eta\to\eta/3$.  The total stress in
Eqn.~\ref{eqn:tstress} then simply reads
\be
\sigmae=G Z + \eta\edot.
\label{eqn:toyStress}
\ee

For any ramp protocol, we set the initial cross sectional area of the
filament $A(0)=1$.  Note that this is in addition to our having set
the initial cylinder length $L(0)=1$ via our choice of length units
above. However it is important to realise that we are not actually
restricting the initial cylinder radius and length to be in any given
ratio: any information about this quantity has simply been lost as a
consequence of our having made the slender filament approximation.

For clarity, we shall drop the  subscript E from the tensile stress
$\sigma_{\rm E}$ for the rest of the paper.

\section{Linear stability analysis}
\label{sec:linear}

We now perform a linear stability analysis to determine the dynamics
of the initial stage in the development of any neck that forms during
the strain-ramp protocol defined above, in which a filament is
strained at a rate $\edotbar$ during times $0<t<t_0$ to a total strain
$\ebar_0=\edotbar t_0$, then held at this strain for all subsequent
times $t>t_0$, during which any viscoelastic stresses that developed
during the the straining process progressively relax.

Our main objective will be to derive fluid universal criteria for the
onset of necking that do not depend on the detailed assumptions of
any given constitutive model. To allow analytical progress, we shall
first perform this calculation within the simplified, generalised
scalar constitutive model defined in Sec.~\ref{sec:protocol}.  The
analogous calculation for the fully tensorial models of
Sec.~\ref{sec:models} is equivalent in principle, but more cumbersome,
and we shall not write it down. Readers are referred to \cite{Hoyle2016a}
for details. Our numerical results below are,
however, for the full tensorial constitutive models. As will be seen,
these (mostly) agree extremely well with our criteria derived in the
simplified scalar model.

In performing the linear stability analysis, we start by considering a
homogeneous ``base state'' corresponding to a filament that remains a
uniform cylinder, with the flow variables uniform along it, both
during the initial straining and the stress relaxation that follows
it.  We shall denote this base state with the subscript $0$.  We then
add to it small amplitude perturbations describing any initial small
heterogeneities along the filament's length, which are the precursor
of a neck.  Expanding the governing equations to first order in the
amplitude of these perturbations yields linearised equations that
govern the dynamics of the perturbations. Our interest then lies in
determining whether these perturbations grow to give a necked state, or
decay to leave a uniform filament. If they do grow, our main aim is to
determine at what stage during the experiment they first start to do so.

Consider first, then, a uniform base state corresponding to a filament
that remains a perfect cylinder during the entire protocol.  This
obeys the homogeneous form of Eqns.~\ref{eqn:1Dmass}
to~\ref{eqn:tforce}, ~\ref{eqn:toy} and~\ref{eqn:toyStress}. The
condition of mass balance accordingly gives
\be
\label{eqn:baseMass}
\dot{A}_0(t)=-\edot_0(t) A_0.
\ee
The tensile force
\be
\label{eqn:baseForce}
F_0(t)=A_0\sigma_0,
\ee
with tensile stress
\be
\label{eqn:baseStress}
\sigma_0(t)=GZ_0+\eta\edot_0.
\ee
The viscoelastic variable evolves according to
\be
\label{eqn:baseZ}
\dot{Z}_0(t)=\edot_0(t)f(Z_0)-\frac{1}{\tau}g(Z_0).
\ee
For the strain ramp protocol of interest here, the strain rate
$\edot_0=\edotbar$ for times $0<t<t_0$, and zero otherwise.

We now add to this homogeneous, time-evolving base state small
amplitude heterogeneous perturbations, which are the precursor of any
neck. As noted above, the dynamics of any necking perturbations under
conditions of a constant Hencky strain rate, indefinitely sustained
for all times $t>0$, have already been studied in our earlier
work~\cite{Hoyle2016a}.  Because such conditions also pertain to the
first (straining) phase of the ramp protocol considered here, our
results from~\cite{Hoyle2016a} automatically apply during that
straining phase.  Accordingly, we shall write here only the equations
governing the fate of any necking perturbations after the straining
has stopped, during the subsequent stress relaxation, when the base
state's strain rate $\edot_0=0$.  For convenience we decompose these
perturbations into Fourier modes with wavevectors $q$ reciprocal to
the distance $z$ along the filament's length:
\begin{equation}
	\begin{pmatrix} \edot(z,t)\\ a(z,t)	\\ Z(z,t)\\
	\end{pmatrix}
= 
	\begin{pmatrix}
		0 \\    a_0	\\		Z_0(t)\\
        \end{pmatrix}
+
	\sum_q\begin{pmatrix}
		\delta\edot(t)\\ \delta a(t)	\\	\delta Z(t)\\
	\end{pmatrix}_q \exp(iqz)
	. \label{eqn:matrixM}
\end{equation}

Although the globally averaged (base state) strain rate $\edot_0=0$
after the straining has stopped, as just noted, flows with zero
$z-$averaged strain rate will nonetheless develop internally along the
filament as part of any necking process, characterised by
$\delta\edot_q(t)$.  The variable $\delta a_q(t)$ determines the
amplitude of any variations in cross sectional area along the
filament, and so characterises the degree of necking at any time $t$.
A key aim in what follows will be to compute the time-dependence of
this quantity for any given fluid and strain ramp protocol. In
particular, we seek to distinguish regimes in which $\delta a_q(t)$
grows in time, leading to the development of a neck, from those in
which it decays, leaving a uniform filament.

We now substitute expression~\ref{eqn:matrixM} into
Eqns.~\ref{eqn:1Dmass} to ~\ref{eqn:tforce}, ~\ref{eqn:toy}
and~\ref{eqn:toyStress}. Expanding these then in successive powers of
the perturbation amplitude, and retaining only terms of first order in
that amplitude, gives a set of linearised equations governing the
dynamics of the perturbations.

The linearised mass balance equation is
\be
\label{eqn:linMass}
\partials{\delta a_q}{t}=-\delta\edot_q.
\ee
The linearised force balance equation is
\be
\label{eqn:linForce2}
0=\sigma_0 \delta a_q + G\delta Z_q + \eta \delta\edot_q,
\ee
and the linearised viscoelastic constitutive dynamics
\be
\partials{\delta Z_q}{t}=\delta \edot_q f(Z_0)+C\delta Z_q,
\ee
in which
\be
\label{eqn:C}
C=-\frac{1}{\tau}g'(Z_0).
\ee
Here prime denotes differentiation with respect to a function's own
argument.

Combining these gives finally 
\begin{equation}
\partial_t
        \begin{pmatrix}
		\delta a(t)	\\ \\	\delta Z(t)\\
	\end{pmatrix}_q 
= \tens{M}(t)
\cdot
        \begin{pmatrix}
		\delta a(t)	\\  \\ 	\delta Z(t)\\
	\end{pmatrix}_q, \label{eqn:2Dset}
\end{equation}
governed by the stability matrix
\be
\tens{M}(t)= 	\begin{pmatrix} \dfrac{\sigma_{\rm 0}}{\eta} & \dfrac{G}{\eta} \\
  & \\
  \dfrac{-f(Z_0)\sigma_{\rm 0}}{\eta}\;\;\;\ & -\dfrac{f(Z_0)G}{\eta} +C \\
  & \\
	\end{pmatrix}.
\label{eqn:MM}
\ee
We note that this stability matrix has inherited the time-dependence
of the relaxing base state conformation and stress variables,
$Z_0(t),\sigma_{\rm 0}(t)$, upon which it depends.  

In writing the necking perturbations in this linearised analysis in
the form $\exp(iqz)$, we have effectively assumed periodic boundary
conditions between the two ends of the filament, thereby implicitly
taking the filament to correspond to a torus being stretched.  In
Ref.~\cite{Hoyle2016a}, we showed that this simplifying assumption
does not strongly affect any conclusions with regards the onset of
necking.  Nonetheless, our full nonlinear simulations in
Sec.~\ref{sec:nonlinear} use more realistic boundary conditions
approximating the no-slip condition that pertains where each end of
the filament meets the rheometer plates. We have checked that the
early time dynamics of our nonlinear simulations, with quasi no-slip,
match those of the linear stability analysis, with periodic boundary
conditions.

We note, however, that the stability matrix $\tens{M}(t)$ does not in
fact depend on the wavevector $q$. In consequence, all Fourier modes
$\exp(iqz)$ are predicted to have the same dynamics.  The mode that
dominates the necking process in practice is therefore expected to be
that which is seeded most strongly by external factors such as thermal
or mechanical noise, endplate effects, or any slight imperfections in
the way the sample is prepared initially. Among these, we expect the
dominant seeding effect to be that imposed by the (quasi) no-slip
condition where the filament meets the rheometer plates, as discussed
in the previous paragraph. This inhibits stretching of the filament in
the vicinity of the sample ends, thereby tending to seed a single neck
mid-sample.

In Eqns.~\ref{eqn:2Dset} and~\ref{eqn:MM}, then, we have arrived at an
equation set governing the linearised dynamics of necking
perturbations, in the regime where the amplitude of these
perturbations remains small. As noted above, the stability matrix
(\ref{eqn:MM}) depends on the time-evolving base state. Our aim now is
to relate any regime in which the degree of necking $\delta a(t)$
grows as a function of time $t$ to characteristic signatures in the
time-evolution of the base state quantities in the stability matrix
(\ref{eqn:MM}). Furthermore, because those quantities by definition
correspond to their counterpart globally measured rheological
quantities in any regime where the necking perturbations remain small,
these signatures in the base state directly correspond to counterpart
signatures in the time-evolving globally measured rheological
quantities.  Our calculation will therefore allow us to report what
signatures in the globally measured rheological quantities correspond
to the development of necking.  Given the correspondence just
discussed, we now drop the 0 subscript from the base state quantities
for simplicity.

Returning to the stability matrix~\ref{eqn:MM}, we note that its
determinant and trace are respectively:
\be
\label{eqn:determinant}
\Delta=\frac{1}{\eta}\,\sigma C=\frac{1}{\eta}\,\sigma\,\frac{\ddot{\sigma}}{\dot{\sigma}}\,,
\ee
and
\be
\label{eqn:trace}
T=-\frac{1}{\eta}\,(f-\sigma).
\ee
(The solvent viscosity $\eta$ is small compared with the zero shear
viscosity of the viscoelastic component, so we ignore any subleading
terms in $\eta$ here and throughout.) In the expression for the
determinant, the second equality follows from the first by combining
Eqn~\ref{eqn:baseZ}, differentiated with respect to time, with
Eqn.~\ref{eqn:baseStress}.  In any regime where $T^2\gg 4|\Delta|$
(which holds unless $|f-\sigma|$ is small), the two eigenvalues of
\ref{eqn:MM} are
\be
\label{eqn:eigenvalue1}
\omega_1=-\frac{1}{f-\sigma}\,\sigma\,\frac{\ddot{\sigma}}{\dot{\sigma}}\,,
\ee
and
\be
\label{eqn:eigenvalue2}
\omega_2=-\frac{1}{\eta}\,(f-\sigma).
\ee

Had these eigenvalues been time-independent, they would have exactly
prescribed the rate of growth (or decay) of necking perturbations.
However, because the base state upon which the eigenvalues depend is
time-dependent, the eigenvalues are themselves time-dependent. In
particular, the first eigenvalue (\ref{eqn:eigenvalue1}) predicts that
the necking perturbations grow (or decay) on a timescale commensurate
with the timescale of the evolution of the base state, and so also of
the evolution of the eigenvalue itself. Put differently: the rate at
which the perturbations are predicted to change itself changes as fast
as the perturbations themselves. To resolve this shortcoming, we
performed a more thorough analysis (not detailed here), to show that
the rate of growth (or decay) of necking associated with this
eigenvalue at any time during the stress relaxation is in fact given
by
\bea
\label{eqn:growth1}
\frac{\dot{\delta a}}{\delta a}&=&\frac{1}{f-\sigma}\,\sigma\left(\frac{\dot{\sigma}}{\sigma}-\frac{\ddot{\sigma}}{\dot{\sigma}}\right),\nonumber\\
 &=& \frac{1}{f-\sigma}\left[-\frac{\sigma^2}{\dot{\sigma}}\right]\,\frac{d^2}{dt^2}\log \sigma.\;\;\textrm{\bf ``stress curvature mode''}\nonumber
\eea
Here and throughout, the notation ``log'' denotes the natural logarithm.

The expression in the square brackets is always positive, because
$\dot{\sigma}<0$ as the stress relaxes over time. In most regimes, the
first term involving $f-\sigma$ is also positive.  Therefore, whether
necking perturbations grow or decay via this mode is determined by the
curvature on a log-linear plot of the stress relaxation function.
Accordingly, we call this mode of necking the ``stress curvature
mode''.

The second eigenvalue (\ref{eqn:eigenvalue2}) is large, predicting a
fast rate of growth (or decay) of necking perturbations on a short
timescale $\eta/G$ compared with the much longer timescale $\tau$ on
which the base state evolves. Concerns about the time-dependence of
this eigenvalue are therefore much less serious: indeed, the
eigenvalue predicts the rate of growth of necking perturbations at any
time during the stress relaxation to excellent approximation:
\bea
\label{eqn:growth2}
\frac{\dot{\delta a}}{\delta a}&=&-\frac{1}{\eta}\,(f-\sigma),\nonumber\\
 &=&-\frac{1}{\eta}\,\frac{1}{A}\,\frac{\partial F}{\partial \epsilon}|_{\rm elastic}.\;\;\;\;\;\;\;\textrm{\bf ``elastic \considere mode''}\nonumber
\eea
The second equality here can be proved by combining
Eqns.~\ref{eqn:baseForce} to~\ref{eqn:baseZ}.  The derivative of the
force $F$ with respect to strain $\epsilon$ that it contains needs
very careful physical interpretation.  Indeed, during the stress
relaxation part of the strain-ramp protocol that is our concern here,
no strain is actually being applied.  The derivative denoted $|_{\rm
  elastic}$ instead defines the incremental change in tensile force
$F$ that {\em would} occur, {\em were} there to be a sudden
incremental strain.  In that very particular sense, this mode provides
the equivalent in these viscoelastic materials of the \considere
criterion for solids~\cite{Considere1885}, which states that necking
will occur in any regime where the force is a decreasing function of
strain. This concept has been discussed in detail in our earlier
papers~\cite{Hoyle2016a,Hoyle2016}.

Our analysis up to this point has been highly general, independent of
the assumptions of any particular constitutive model.  We now seek to
apply the two necking criteria (``stress curvature'' and ``elastic
\considere'') that it has allowed us to derive to the various
tensorial constitutive models of Sec.~\ref{sec:models}.  As a first
step, we recall from Ref.~\cite{Hoyle2016a} that the dynamics of our
simplified scalar model of Eqn.~\ref{eqn:toy} can be made closely to
mimic the Oldroyd B, Giesekus and non-stretch Rolie-Poly models by
taking
\be
f=3+n Z-\tfrac{2}{3}(1+\beta) Z^2.
\ee
and
\be
g=Z+\alpha Z^2.
\ee
The case $\alpha=0, \beta=-1, n=2$ gives a good approximation to the
Oldroyd B model; $0<\alpha< 1, \beta=-1, n=2$ to the Giesekus model;
and $\alpha=0, 0<\beta<1, n=1$ to the non-stretch Rolie-Poly model. We
shall use these simplified scalar forms of the three tensorial
constitutive models in the next three subsections in turn to
understand our numerical results for each of these three models. We
emphasise that those numerical results were obtained in the full
tensorial form of each model.

Recognising that during the stress relaxation the strain rate
$\edotbar=0$, and therefore that the total tensile stress $\sigma$ and
the viscoelastic strain variable $Z$ coincide up to a prefactor $G=1$,
we shall use the symbols $Z$ and $\sigma$ interchangeably in what
follows.

\subsection{Oldroyd B model}

\begin{figure*}
\includegraphics[width=1.0\textwidth]{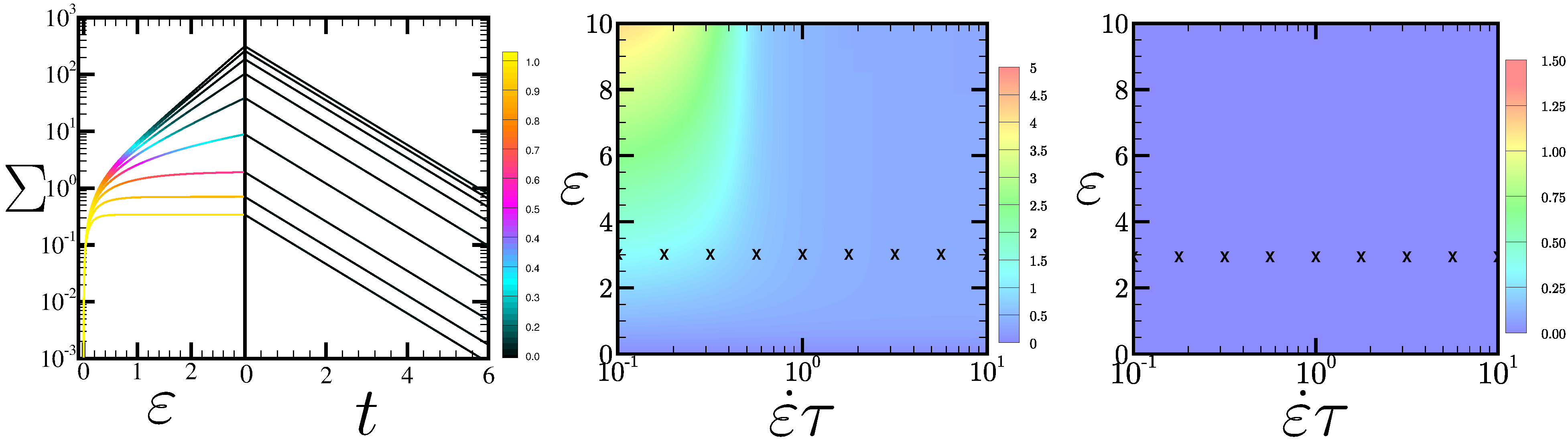}
\caption{Linearised necking dynamics in the Oldroyd B model during and
  after extensional straining. {\bf Left:} stress as a function of
  strain during straining, and as a function of time after the
  straining has stopped. Colour scale here shows the instantaneous
  rate of necking per unit strain, $\delta a'/\delta a$, at any strain
  during straining (where $'$ denotes differential with respect to
  strain), and the instantaneous rate of necking per unit time,
  $\dot{\delta a}/\delta a$ at any time after straining.  {\bf
    Centre:} colour map (with a logarithmic colour scale) of $\delta
  a(\epsilon,\edot)/\delta a(\epsilon=0,\edot)$, showing the total
  degree of necking that accumulates during the initial straining
  process, normalised by its value before straining commenced, as a
  function of the strain imposed and the rate at which it is imposed.
  {\bf Right:} colour map (with a logarithmic colour scale) of $\delta
  a(\epsilon,\edot,t\to\infty)/\delta a(\epsilon,\edot,t=0)$, showing
  the total degree of necking that accumulates during the full stress
  relaxation process after the strain has stopped, normalised by its
  value at the start of the stress relaxation (i.e., by its value at
  the end of the straining process), as a function of the total strain
  that had initially been applied, and the strain rate at which it had
  been applied. Parameter values: $\eta=10^{-4}$.}
\label{fig:OldroydB} 		
\end{figure*}

Our numerical results for the linearised necking dynamics of the full
tensorial Oldroyd B model are shown in Fig.~\ref{fig:OldroydB}. The
left panel shows the evolution of the stress: first (in the left
subpanel of this left panel) as a function of strain while the sample
is being stretched, and then (in the right subpanel) as a function of
time during the relaxation of the stress after the straining has
stopped.  The colourscale superposed on this stress evolution
indicates the rate of growth of necking perturbations at any stage in
the protocol. (Any regime where the growth is negative, indicating
decay, is shown in black.) As can be seen, no growth of necking occurs
during the stress relaxation after the straining stops.

This can be understood by applying the two criteria developed in
Sec.~\ref{sec:linear} to the simplified scalar version of the Oldroyd
B model. In this, the loading function $f=3+2\sigma$. This gives
$f-\sigma=3+\sigma$, which is is always positive, both during the
initial straining and after the straining stops.  Accordingly, the
elastic \considere mode remains stable at all times and causes no
necking. The relaxation function $g=\sigma$, so after the straining
has stopped the stress relaxes as $\dot{\sigma}=-\sigma/\tau$, and
therefore as $\sigma(t)=\sigma(0)\exp(-t)$.  (We define the origin of
time here to coincide with the end of straining. Previously we used it
to indicate the start of straining.)  The curvature of the stress
decay on a log-linear plot is therefore zero, and the stress curvature
mode predicts neutral stability with respect to necking. In this way,
our analytically derived necking criteria, applied to the (simplified,
scalarised) Oldroyd B model, predict no growth of necking
perturbations during the stress relaxation post-straining, consistent
with our numerical results in Fig.~\ref{fig:OldroydB}.

Some necking does however occur during the first part of the protocol,
while the strain is being applied. As noted above, during this
  regime the flow coincides with the simpler protocol in which a
constant Hencky strain rate is applied indefinitely for all times
after the initial switch on. It is accordingly covered by our results
in Ref.~\cite{Fielding2011,Hoyle2016a}, to which the reader is
referred for details.  Here, we merely recall that in the Oldroyd B
model any significant necking during straining arises only for strain
rates $\edotbar<0.5$, and via the stress curvature mode (generalised
to the case of a non-zero strain rate; the stress curvature mode
derived above holds for the particular case of zero strain rate).

The results just discussed in the left panel of
Fig.~\ref{fig:OldroydB} pertain to nine different strain ramp
protocols, each with a different value of the imposed strain rate
$\edotbar$, but each with the same total applied strain $\ebar_0=3.0$,
as indicated by the crosses in the middle panel of the figure.  Beyond
these nine individual runs, we also investigated a much fuller range
of values of $\edotbar$ and $\ebar$. The results are represented in a
compact way in the middle and right panels of Fig.~\ref{fig:OldroydB}.

In the middle panel, the colourscale at any coordinate pair
$(\edotbar,\ebar_0)$ denotes the total amount of necking that is
predicted (at the level of this linear calculation) to accumulate
during the first, straining part of a strain ramp protocol, in which a
total strain of $\ebar_0$ is applied at a constant rate $\edotbar$.
(In this way, the colourscale at the location of the nine crosses in
the middle panel is effectively an integral over the data shown by the
colourscale in the individual runs of the left subpanel of the left
panel.)

In the right panel, the colourscale at each coordinate pair
$(\edotbar,\ebar_0)$ denotes the total amount of necking that further
accumulates during the stress relaxation after the straining has
stopped, for a ramp in which a total strain of $\ebar_0$ has been
applied at a constant rate $\edotbar$.  (In this way, the colourscale
at the location of the nine crosses in the right panel is effectively
an integral over the data shown by the colourscale in the individual
runs of the right subpanel of the left panel.) This panel is
essentially however redundant in this particular model, given that no
further necking is predicted after the straining stops.

To summarise, in the Oldroyd B model some necking occurs during the
straining process itself for imposed strain rates $\edotbar<0.5$. No
further necking is predicted to accumulate during the stress
relaxation after the straining stops.

\begin{figure*}
\includegraphics[width=1.0\textwidth]{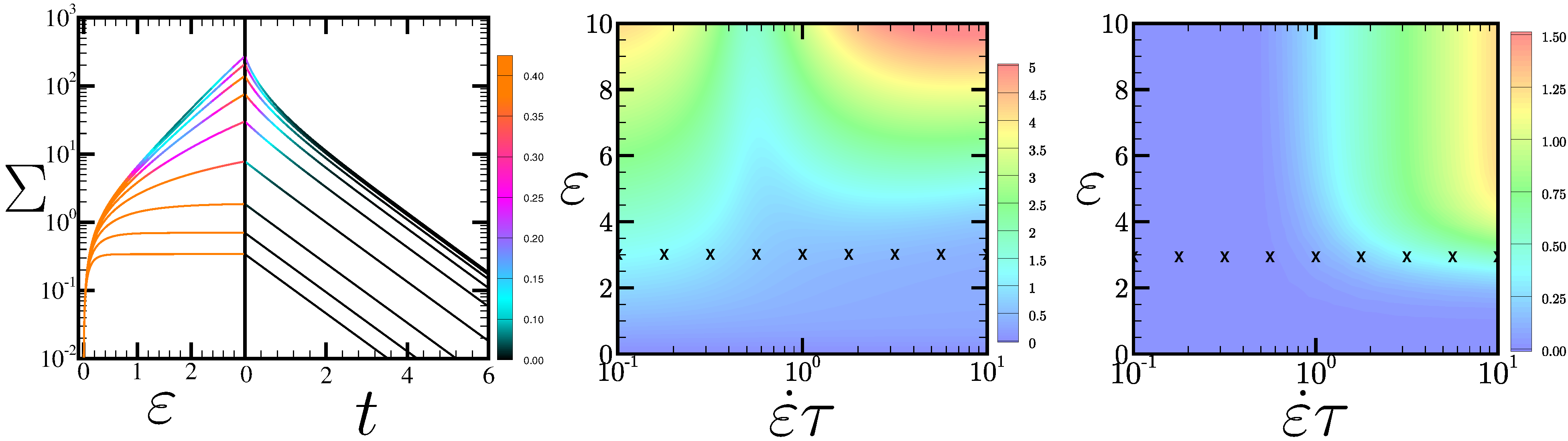}
\caption{Linearised necking dynamics in the Giesekus model during and
  after extensional straining. {\bf Left:} stress as a function of
  strain during straining, and as a function of time after the
  straining has stopped. Colour scale here shows the instantaneous
  rate of necking per unit strain, $\delta a'/\delta a$, at any strain
  during straining (where $'$ denotes differential with respect to
  strain), and the instantaneous rate of necking per unit time,
  $\dot{\delta a}/\delta a$ at any time after straining.  {\bf
    Centre:} colour map (with a logarithmic colour scale) of $\delta
  a(\epsilon,\edot)/\delta a(\epsilon=0,\edot)$, showing the total
  degree of necking that accumulates during the initial straining
  process, normalised by its value before straining commenced, as a
  function of the strain imposed and the rate at which it is imposed.
  {\bf Right:} colour map (with a logarithmic colour scale) of $\delta
  a(\epsilon,\edot,t\to\infty)/\delta a(\epsilon,\edot,t=0)$, showing
  the total degree of necking that accumulates during the full stress
  relaxation process after the strain has stopped, normalised by its
  value at the start of the stress relaxation (i.e., by its value at
  the end of the straining process), as a function of the total strain
  that had initially been applied, and the strain rate at which it had
  been applied.  Parameter values: $\eta=10^{-4},\alpha=0.01$.}
\label{fig:Giesekus} 		
\end{figure*}

\subsection{Giesekus model}
\label{sec:Giesekus}

Our numerical results for the linearised necking dynamics of the full
tensorial Giesekus model are shown in Fig.~\ref{fig:Giesekus}. These
are presented in the same way as for the Oldroyd B model in
Fig.~\ref{fig:OldroydB} so we shall not explain the figure in detail
again, but focus on drawing out the main similarities and differences
between the two models.

In the left panel of Fig.~\ref{fig:Giesekus}, we see that the
relaxation of the stress post-straining displays some initial upward
curvature (in the log-linear representation used here):
\be
\frac{d^2}{dt^2}\log \sigma>0.
\ee
Associated with this upward curvature is some instability to necking
at early times in this stress relaxation process, as seen by the
bright patch in the colourscheme. This then gradually diminishes over
time.

This behaviour can be understood within the simplified scalar form of
the Giesekus model as follows.  Its relaxation function
$g=\sigma+\alpha\sigma^2$ (with Oldroyd B dynamics recovered only for
$\alpha=0$).  Accordingly, during the stress relaxation after the
straining has stopped, $\dot\sigma/\sigma$ is more negative for larger
values of sigma: the stress relaxation proceeds more quickly at
earlier times, then progressively slows down as the stress decays.
This indeed gives upward curvature of the rate of decay of the
log-stress, as seen in the right subpanel of the left panel of
Fig.~\ref{fig:Giesekus}.  Accordingly, the stress curvature mode
predicts instability to necking during the first part of the stress
decay post-straining. (In contrast, the loading function of the scalar
Giesekus model is the same as that of scalar Oldroyd B: $f=3+2\sigma$.
Accordingly, the elastic \considere mode remains stable in the
Giesekus model, as in Oldroyd B, both during and after straining.)

The results in the left panel of Fig.~\ref{fig:Giesekus} pertain to
nine different strain ramps, each performed at a different value of
the strain rate $\edotbar$, but each with a total applied strain
$\ebar_0=3.0$, as indicated by the crosses in the middle panel. 

In the middle and right panels of Fig.~\ref{fig:Giesekus} we explore a
much fuller range of pairs of values of $\edotbar,\ebar_0$. As in
Fig.~\ref{fig:OldroydB} for Oldroyd B, the colourscale in the middle
panel shows the total degree of necking that accumulates during the
straining process for a ramp performed with any given pairing of
$\edotbar,\ebar_0$.  As noted above, during this straining part of the
protocol the dynamics will coincide with those of the simpler protocol
in which a constant strain rate is applied indefinitely for all times
after the initial switch on. Accordingly, our results in the middle
panel for the degree of necking that accumulates during straining
follow directly from those in Fig.~3b of our earlier work of
Ref.~\cite{Hoyle2016a} for that protocol of a constant applied strain
rate. In the right panel of Fig.~\ref{fig:Giesekus}, we show by
  the colourscale the total degree of necking that further accumulates
  during the stress relaxation after the straining stops, for a ramp
  that had been performed at any $\edotbar,\ebar_0$.

Compared with Oldroyd B, the most significant feature (at least of
principle) in the Giesekus model is that some further necking does
take place during the stress relaxation after the straining stops, for
ramps with rates $\edotbar\gtrsim 1$ and strains $\ebar_0\gtrsim 1$,
via the stress curvature mode.  Comparing the colourscales in the
middle and the right panels, however, we see that for any given ramp
as specified by $\edotbar,\ebar_0$, the degree of necking that further
accumulates after the straining stops (right panel) is always very
modest compared with that which took place during the straining itself
(middle panel).  Indeed, for any case in which noticeable further
necking might in principle have accumulated post-strain, the sample
will almost certainly have anyway already failed entirely during the
straining process itself.

Our analytical calculations in this section were performed in the
simplified scalar version of the Giesekus model. In contrast, our
numerical results in Fig.~\ref{fig:Giesekus} are for the fully
tensorial form of the model. To demonstrate the equivalence of these,
Fig.~\ref{fig:toy} in the Appendix shows numerical results for the
scalar version of the Giesekus model.  Reassuringly, we see close
agreement with those of Fig.~\ref{fig:Giesekus} for the fully
tensorial form.  Further evidence of the close correspondence between
the scalar version of each model (Oldroyd B, Giesekus, non-stretch
Rolie-Poly) and its full tensorial counterpart is given in
Ref.~\cite{Hoyle2016a}.

\begin{figure*}
\includegraphics[width=1.0\textwidth]{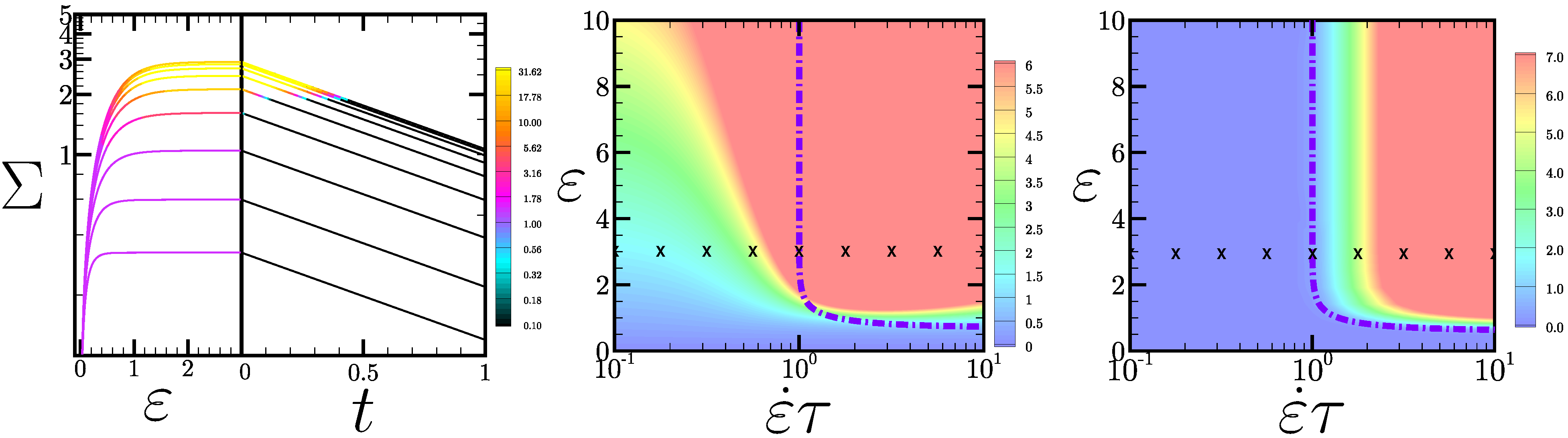}
\caption{Linearised necking dynamics in non-stretching Rolie Poly
  model during and after extensional straining.  {\bf Left:} stress as
  a function of strain during straining, and as a function of time
  after the straining has stopped. Colour scale here shows the
  instantaneous rate of necking per unit strain, $\delta a'/\delta a$,
  at any strain during straining (where $'$ denotes differential with
  respect to strain), and the instantaneous rate of necking per unit
  time, $\dot{\delta a}/\delta a$ at any time after straining.  {\bf
    Centre:} colour map (with a logarithmic colour scale) of $\delta
  a(\epsilon,\edot)/\delta a(\epsilon=0,\edot)$, showing the total
  degree of necking that accumulates during the initial straining
  process, normalised by its value before straining commenced, as a
  function of the strain imposed and the rate at which it is imposed.
  {\bf Right:} colour map (with a logarithmic colour scale) of $\delta
  a(\epsilon,\edot,t\to\infty)/\delta a(\epsilon,\edot,t=0)$, showing
  the total degree of necking that accumulates during the full stress
  relaxation process after the strain has stopped, normalised by its
  value at the start of the stress relaxation (i.e., by its value at
  the end of the straining process), as a function of the total strain
  that had initially been applied, and the strain rate at which it had
  been applied.  Parameter values: $\eta=0.0033$, $\beta=0.0$.}
\label{fig:nRP} 		
\end{figure*}

\subsection{Non-stretch Rolie-Poly model}

Our numerical results for the linearised necking dynamics of the
tensorial Rolie-Poly model without chain stretch (which we shall
  call the nRP model) are shown in Fig.~\ref{fig:nRP}, in the same
format as in Figs.~\ref{fig:OldroydB} and~\ref{fig:Giesekus} for the
Oldroyd B and Giesekus models.  As can be seen in the left panel,
after the straining stops the stress relaxes exponentially.  During
the first part of this stress relaxation, the system displays
instability to necking for those ramps in which the stress had
exceeded a threshold value $\sigma_{\rm c}$ approximately equal to two
by the end of the straining process.

These observations can be understood within the simplified scalar
version of the nRP model as follows. The relaxation function of this
model is the same as for the (scalar) Oldroyd B model: $g=\sigma$.
This gives exponential stress relaxation $\sigma=\exp(-t)$ as a
function of the time $t$ since the straining stops, consistent with
the numerics: the stress decay plotted in log-linear representation is
linear. In this way, the stress curvature mode is always stable
against necking post-strain. In contrast, the loading function
$f=3+\sigma-\tfrac{2}{3}(1+\beta)\sigma^2$, and the factor $f-\sigma$
that appears in the criterion for the elastic \considere mode to be
unstable will become negative if sufficient stress $\sigma >
\sigma_{\rm c} = 3/\sqrt{2}\approx 2.12$ (for $\beta=0.0$) develops
during the straining process.  Necking instability will then arise via
this elastic \considere mode, and persist after straining stops until
such a time as the stress decays to be again below $\sigma_{\rm c}$,
returning $f-\sigma$ to positivity.  This can be seen via the bright
patch in the colourscale during the early part of the stress
relaxation in the left panel of Fig.~\ref{fig:nRP}, for the ramps
performed at the higher strain rate values.

In this left panel, we show results for strain ramps performed at nine
different values of the imposed strain rate $\edotbar$, with a total
imposed strain $\ebar_0=3.0$ in each case, as indicated by the
crosses in the middle panel.  We now consider more broadly in what
region of the full plane of values of $\edotbar,\ebar_0$ will
$f-\sigma$ become negative by the end of the straining process, and so
by the start of the stress relaxation post-ramp, signifying
instability to necking via the elastic \considere mode during the
first part of the stress relaxation in the manner just described. 

We studied this by integrating the model equations to obtain the base
state stress $\sigma(\ebar,\edotbar)$ as a function of accumulated
strain $\ebar$ and imposed strain rate $\edotbar$.  Doing so, we find
$f-\sigma$ to be negative at the end of the straining process, and so
at the start of the stress relaxation, for values of $\ebar,\edotbar$
above and to the right of the dot-dashed line in the middle and right
panels of Fig.~\ref{fig:nRP}.  In the limit of large strain rates
$\edotbar\to\infty$, the criterion for $f-\sigma$ to be negative tends
to the condition
\be
\ebar>-\frac{1}{3}\log\left(\frac{\sqrt{2}-1}{2+\sqrt{2}}\right)\approx
0.703, 
\ee 
as seen by the asymptote of the dot-dashed line at the the right hand
side of these panels.  In the limit of large strains $\ebar\to\infty$,
the criterion for $f-\sigma$ to be negative tends to the condition
$\edotbar>1$, as seen by the asymptote at the top of the panels.  

The right panel of Fig.~\ref{fig:nRP} shows as a colourscale the total
degree of necking predicted to accumulate during the entire stress
relaxation after the straining has stopped as a result of this elastic
\considere mode, as a function of the rate $\edotbar$ at which the
strain had been applied, and the total strain imposed, $\ebar_0$.
Consistent with the above analytical prediction, significant necking
indeed occurs during the stress relaxation in the region of the plane
of $\edotbar,\ebar_0$ above the dot-dashed line.

This finding however proves to be mainly of pedagogical interest in
this non-stretching version of the model, for the following reason.
Comparing the right panel of Fig.~\ref{fig:nRP} (showing the total
necking that accumulates post-strain) with the middle panel, which
shows the total necking accumulated during the straining process
itself, we see that for any given pair of values of $\edotbar,\ebar_0$
in which significant further necking might in principle accumulate
after the straining stops, the sample will almost certainly have
anyway failed altogether due to the very large degree of necking
predicted during the straining process.

To summarise our results so far: in this and the previous two
subsections, we have discussed the linearised necking dynamics of the
Oldroyd B, Giesekus and non-stretch Rolie-Poly models during the
stress relaxation after the straining stops. In the Oldroyd B model,
we saw no further necking during the stress relaxation (beyond any
that had already accumulated during straining in ramps for which
$\edotbar<0.5$).  In contrast, the Giesekus and non-stretch Rolie-Poly
models do predict some further necking during the stress relaxation. In
each case, however, a much larger degree of necking is predicted to
have already occurred during the straining process itself, likely
causing the filament fail altogether during the straining.  The
discussion of the previous three subsections can therefore be viewed
as being primarily pedagogical, in helping us to understand the
general conditions under which necking might occur after straining
stops.  Overall, perhaps the principal physical conclusion for the
dilute or semi-dilute solutions modelled by the Oldroyd B and Giesekus
models is that necking after straining stops is unlikely to be an
important physical effect compared with any that takes place during
the straining process itself.

In the next subsection, we turn to the Rolie-Poly model of
concentrated solutions and melts of entangled linear polymers, and
wormlike micellar surfactant solutions, with chain stretch included.
In this case, we shall find an important new regime in which necking
is suppressed during the straining process itself, due to the
accumulation of chain stretch, but in which significant necking then
occurs with a delayed onset after the straining stops, as the chain
stretch relaxes on the short timescale $\taus$, but before the
orientational contribution to the stress relaxes on the much longer
timescale $\taud$.  Such a physical scenario was predicted in the
insightful earlier work of the Copenhagen group in this context of
extensional necking~\cite{Lyhne2009,Rasmussen2011}. (An analogous
scenario was predicted by one of the present authors, with others, in
the context of shear banding following a shear strain
ramp~\cite{Moorcroft2013,Adams2009}. We shall return to discuss this in
Sec.~\ref{sec:banding} below.) Here we build on that intuition, in
particular by providing an analytical criterion for necking, and
showing that it agrees fully with the regimes and rates of necking
seen in the numerical simulations.

\subsection{Stretching Rolie-Poly model}

\begin{figure*}
\includegraphics[width=1.0\textwidth]{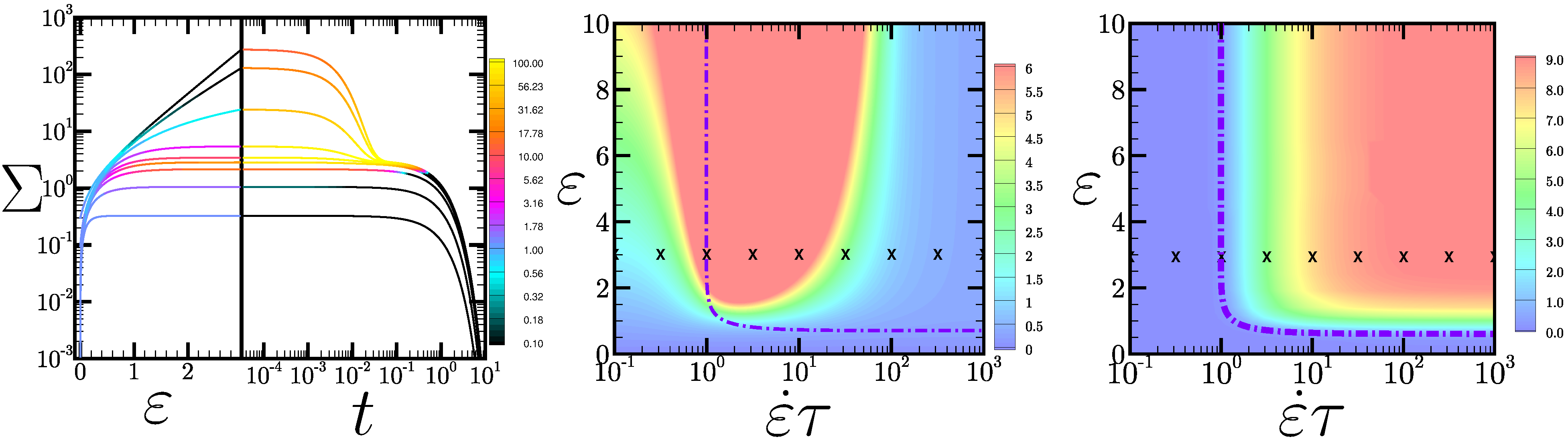}
\caption{Linearised necking dynamics in the stretching Rolie Poly
  model during and after extensional straining.  {\bf Left:} stress as
  a function of strain during straining, and as a function of time
  after the straining has stopped. Colour scale here shows the
  instantaneous rate of necking per unit strain, $\delta a'/\delta a$,
  at any strain during straining (where $'$ denotes differential with
  respect to strain), and the instantaneous rate of necking per unit
  time, $\dot{\delta a}/\delta a$ at any time after straining.  {\bf
    Centre:} colour map (with a logarithmic colour scale) of $\delta
  a(\epsilon,\edot)/\delta a(\epsilon=0,\edot)$, showing the total
  degree of necking that accumulates during the initial straining
  process, normalised by its value before straining commenced, as a
  function of the strain imposed and the rate at which it is imposed.
  {\bf Right:} colour map (with a logarithmic colour scale) of $\delta
  a(\epsilon,\edot,t\to\infty)/\delta a(\epsilon,\edot,t=0)$, showing
  the total degree of necking that accumulates during the full stress
  relaxation process after the strain has stopped, normalised by its
  value at the start of the stress relaxation (i.e., by its value at
  the end of the straining process), as a function of the total strain
  that had initially been applied, and the strain rate at which it had
  been applied.  Parameter values: $\eta=10^{-4},\beta=0.0$ and
  $\taur=0.0083$ (corresponding to an entanglement number $Z_{ent}=40$).  }
\label{fig:sRP0} 
\end{figure*}

Our numerical results for the Rolie-Poly model with chain stretch
included are shown in Fig.~\ref{fig:sRP0}, in the same format as in
Figs.~\ref{fig:OldroydB} to~\ref{fig:nRP} for the Oldroyd B, Giesekus
and non-stretch Rolie-Poly models. The new necking physics inherent to
this model with chain stretch included is evident already in the left
panel of Fig.~\ref{fig:sRP0}, which shows the stress as a function of
strain during the straining process, and as a function of time during
the relaxation after the straining has stopped. The superposed
colourscale indicates the rate of necking at any given strain (or
time).

For the lowest imposed strain rates, $\edotbar\lesssim 1/\taud=1$,
there is a modest rate of necking during the straining process itself,
then essentially no further necking during the stress relaxation after
the straining stops. For intermediate strain rates,
$1/\taud\lesssim\edotbar\lesssim 1/\taus$, there is a high rate of
necking during the straining. This originates from a significant
vestige, in this stretching form of the model, of the elastic
\considere mode of necking instability in the non-stretching model.
(We shall return to discuss this in more detail below.)  Fast necking
is also predicted to continue after the straining stops (for the same
reason).  However that prediction of post-strain instability may be
largely irrelevant: the very fast necking during the straining process
itself is likely to cause the filament to fail altogether even before
the straining stops.


The important new physics of the stretching model arises at high
imposed strain rates, $\edotbar\gtrsim 1/\taus$.  In this regime, the
sample is strongly stabilised against necking during the straining
process by the accumulation of chain stretch.  (This was discussed at
length in our earlier work~\cite{Fielding2011,Hoyle2016a}, and we do
not repeat the details here.) Once the straining stops and the stress
starts to relax, however, strong necking instability sets in.

Two separate regimes are evident in the stress relaxation. The first
occurs on the fast timescale $\taus$ over which the chain stretch
relaxes, with the stress quickly falling from its high initial value
to an intermediate plateau value.  During this first regime we see a
fast rate of necking, which in fact becomes ever faster as that first
regime proceeds to completion.  The second regime occurs on the much
slower timescale $\taud$ of reptation, with the stress finally
decaying from its intermediate plateau value towards zero.  The rate
of necking falls in tandem with the stress, and the system is
predicted to recover stability against necking once the stress falls
below a threshold value $\sigma_{\rm c}\approx 2$.  This predicted
return to stability may be unimportant in practice, however, as the
sample is likely to have failed altogether by this time.

The left panel of Fig.~\ref{fig:sRP0} pertains to nine different
strain ramps, each performed at a different imposed strain rate
$\edotbar$, but with a total imposed strain $\ebar_0=3.0$ in each
case, as indicated by the crosses in the middle panel. In the middle
and right panels we explore a much wider range of pairs of values of
$\ebar_0,\edotbar$.  As before, the colourscale in the middle panel
shows the total degree of necking that accumulates during the
straining process itself, for any given imposed $\ebar_0,\edotbar$.
The right panel shows the total degree of necking predicted further to
accumulate during the stress relaxation post-strain, again as a
function of the variables $\ebar_0,\edotbar$ that prescribed the ramp.

The important new physical regime in this stretching version of the
Rolie-Poly model is that for which the imposed strain rate
$\edotbar\gtrsim 1/\taus$ and the imposed strain $\ebar_0\gtrsim
0.703$. For such ramps, the accumulation of chain stretch stabilises
the filament against necking during the straining process itself.  A
strong delayed necking then sets in as the chain stretch relaxes
post-straining. For imposed strains $\ebar_0<0.703$ (recall that
$\ebar_0=0.703$ is the asymptote of the dot-dashed line at high strain
rates), no significant necking occurs either during or after
straining. For value pairings above the dot-dashed line but with
$\edotbar<1/\taus$, the sample is likely to fail altogether during the
straining process itself, with any prediction of further necking
post-ramp accordingly essentially irrelevant.

We now sketch an analytical calculation to enable us to understand
these numerical results of Fig.~\ref{fig:sRP0} in more detail.  As
ever, at the level of a slender filament approximation the condition
of mass balance gives:
\be
\partials{A}{t} + V\partials{A}{z}  = -\dot\varepsilon A. 
\label{eqn:1Dmassb} 
\ee
The force balance condition gives
\be
0 = \partials{F}{z},
 \label{eqn:1Dmomb} 
\ee
in which the tensile force
\be
\label{eqn:tforceb}
F(t)=A(z,t)\sigmae(z,t),
\ee
and the total tensile stress 
\be
\label{eqn:tstressb}
\sigmae = G\left(W_{zz} - W_{xx} \right) + 3\eta\dot\varepsilon.  \ee

The evolution in flow of the components $W_{zz}$ and $W_{xx}$ is given
by Eqn.~\ref{eqn:sRP}. For simplicity in this analytical calculation
we set the convective constraint release parameter $\beta=0$, as in
our numerical calculations of Fig.~\ref{fig:sRP0}. (Additional
numerical results shown in Fig.~\ref{fig:sRP1} in the Appendix confirm
essentially the same physical scenario for $\beta=1$, at the opposite
end of the allowed range $0<\beta<1$.)  We adopt the notation for the
viscoelastic tensile stress variable $\sigma=W_{zz}-W_{xx}$ (recall
that the modulus $G=1$ in our units) and for the chain stretch
variable $T=\sum_i W_{ii}$, and then further define the variable
$s=3\sigma/T$ that quantifies the chain orientation. The components of
Eqn.~\ref{eqn:sRP} then yield equations of motion respectively for the
variable $s$ that quantifies the polymer chain orientation, and for
the variable $T$ that quantifies polymer chain stretch:
\bea
\frac{Ds}{Dt}&=&\edot f(s)-\frac{1}{\taud}\frac{3s}{T},\nonumber\\
\frac{DT}{Dt}&=&\frac{2}{3}\edot sT-\frac{1}{\taud}(T-3)-\frac{2}{\taus}\left(1-\sqrt{\frac{3}{T}}\right)T.\nonumber
\\
\eea
in which (for $\beta=0$) 
\be
f(s)=3+s-\frac{2}{3}s^2.
\ee

With these equations, we now perform the usual linear stability
analysis, considering an underlying homogeneous base state
corresponding to a cylinder that remains perfectly uniform, to which
are then added small amplitude spatially varying perturbations, which
are the precursor of a neck. We specialise to the time regime after
the straining has stopped, recalling that the behaviour during
straining has already been studied in Ref.~\cite{Hoyle2016a}.

The quantities pertaining to the homogeneous base state relax
post-strain according to
\bea
\dot{s}&=&-\frac{1}{\taud}\frac{3s}{T},\nonumber\\
\dot{T}&=&-\frac{1}{\taud}(T-3)-\frac{2}{\taus}\left(1-\sqrt{\frac{3}{T}}\right)T.
\label{eqn:sRPbase}
\eea
(We have omitted the subscript 0 used previously to denote the base
state, having already noted above that the base state quantities
coincide with their experimentally measured global rheological
counterparts as long as the necking perturbations remain small.) From
the second of Eqns.~\ref{eqn:sRPbase} one can show that the chain
stretch relaxes back to its equilibrium value $T=3$ as a function of
the time $t$ post-strain on a fast timescale $\taus\ll\taud$ as
\be
\label{eqn:stretch}
\sqrt{T(t)}=\sqrt{T(0)}+(\sqrt{3}-\sqrt{T(0)})\left[1-\exp\left(-\frac{t}{\taus}\right)\right].
\ee
Note that for ramps with strain $\ebar_0\gtrsim 1 $ imposed at a fast
rate $\edotbar\gg 1$, the initial value $T(0)$ at the start of the
stress relaxation (ie, at the end of the straining) can be large.

The variable $s$ associated with chain orientation relaxes on the much
longer timescale $\taud\equiv 1$ according to
\be
\label{eqn:orient}
s(t)=s(0)\exp(-t).
\ee

Taken together, Eqns.~\ref{eqn:stretch} and~\ref{eqn:orient} confirm
that the tensile stress $\sigma=sT/3$ has two regimes of relaxation:
the first associated with relaxation of the chain stretch variable $T$
on the fast timescale $\taus$, and the second with relaxation of the
chain orientation variable $s$ on the slower timescale $\taud$,
consistent with our numerical results in the left panel of
Fig.~\ref{fig:sRP0}.

Turning now to consider the small amplitude necking perturbations to
the homogeneous underlying base state just discussed, we write
linearised equations analogous to those in Eqns.~\ref{eqn:linMass}
to~\ref{eqn:MM} for the simplified scalar model. From these, it is
possible to show (though we do not provide the details) that the
rate of necking at any time during the stress relaxation process is
given by
\bea
\frac{\dot{\delta a}}{\delta a}&=&-\frac{1}{3}\left(f(s)-s\right)\left(\frac{\dot{T}}{T}-\frac{\ddot{T}}{\dot{T}}\right),\nonumber\\
&=&-\frac{1}{3}\left(f(s)-s\right)\left[-\frac{T}{\dot{T}}\right]\frac{d^2}{dt^2}\log T.
\label{eqn:sRPneck}
\eea
(The quantity in the square brackets is always positive, because the
trace $T$ decays as a function of time post-straining, $\dot{T}<0$.)
This result contains two important factors that closely mirror
analogous quantities in our earlier criteria for necking in the
simplified scalar constitutive model. The first resembles the factor
$f(\sigma)-\sigma$ in the criterion for instability of the elastic
\considere mode, but expressed now in terms of the chain orientation
variable $s=3\sigma/T$. (In the non-stretch Rolie-Poly model, $T=3$ at
all times and $s=\sigma$.) The second resembles the factor
$\tfrac{d^2}{dt^2}\log \sigma$ of the stress curvature mode, but
expressed in terms of the chain stretch variable $T=\sum_i T_{ii}$.

We consider now the behaviour of each of these two factors in turn, as
a function of the time $t$ after the straining stops. It is easy to
show from the second of Eqns.~\ref{eqn:sRPbase} that the factor
involving $T$ as written in its form in the first of
Eqns.~\ref{eqn:sRPneck} obeys (for $\taud\gg \taus$) the relation
\be
\label{eqn:Tcurve}
\frac{\dot T}{T}-\frac{\ddot{T}}{\dot{T}}=\frac{1}{\taus}\sqrt{\frac{3}{T}}.
\ee
This is always positive, and is in general large in magnitude because
the stretch relaxation timescale $\taus$ is small.  Having thus shown
that the $T$-curvature factor in Eqn.~\ref{eqn:sRPneck} is always
positive, we recognise that any counterpart of the curvature mode of
instability of the scalar model will always be stable in this context
of the stretching Rolie-Poly model.

In consequence, the stability/instability to necking must be
determined by the other, elastic \considere-like factor $f(s)-s$ in
Eqn.~\ref{eqn:sRPneck}, with instability for $f-s<0$. For high imposed
strain rates, this orientational variable $s$ in the stretching
Rolie-Poly model follows essentially the same dynamics as the variable
$\sigma$ in the non-stretching form of the model.  Accordingly, the
factor $f(s)-s$ can become negative during strain ramps of
sufficiently large strain performed sufficiently quickly.  The region
of the plane of imposed strain $\ebar_0$ and strain rate $\edotbar$
for which $f(s)-s$ is indeed negative by the end of the straining
process is that above the dot-dashed line in the middle and right
panels of Fig.~\ref{fig:sRP0}.  This coincides with the counterpart
line shown in Fig.~\ref{fig:nRP} for the non-stretching model, and
asymptotes to $\ebar_0=0.703$ as $\edotbar\to\infty$.

In any regime where indeed $f-s<0$, Eqn.~\ref{eqn:Tcurve} substituted
into Eqn.~\ref{eqn:sRPneck} tells us that the necking will develop
with a characteristic rate that scales as $1/\taus$, and that actually
becomes progressively larger during the course of the first regime of
stress relaxation post-straining, as the chain stretch relaxes from
its large initial value $T(0)$ to its equilibrium value $T=3$.

\begin{figure}
\includegraphics[width=0.49\textwidth]{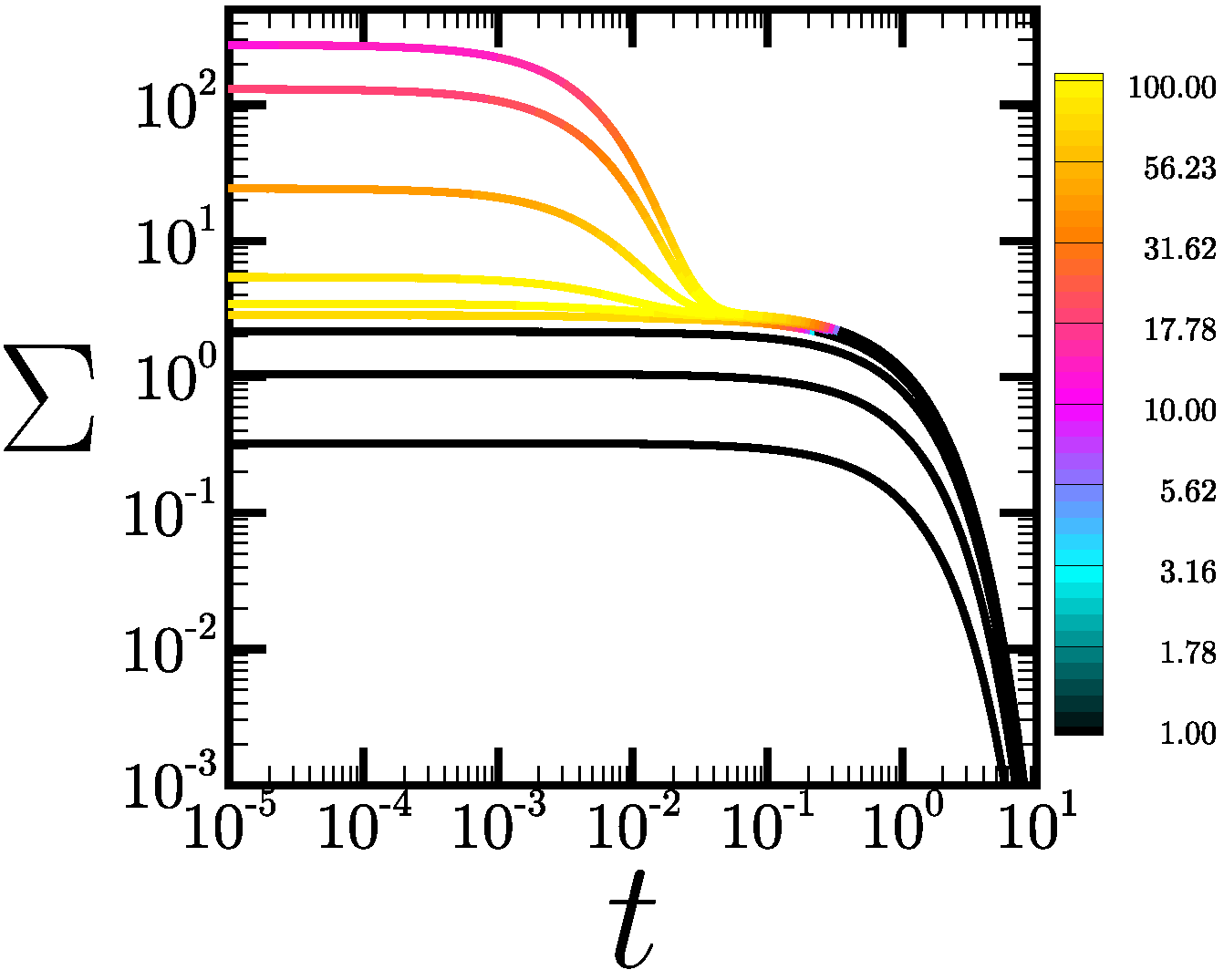}
\caption{
  Relaxation of the stress $\sigma=sT/3$, with the colourscale showing
  the analytical prediction~\ref{eqn:sRPneck} for the rate of growth
  of necking perturbations. This analytical prediction for the rate of
  growth of necking should be compared with the corresponding
  numerical results shown by the colourscale in the right subpanel for
  the left panel of Fig.~\ref{fig:sRP0}.}
\label{fig:analytics} 		
\end{figure}

The rate of growth of necking during the stress relaxation, as
predicted by the analytical form in Eqn.~\ref{eqn:sRPneck}, is plotted
as a colourscale in Fig.~\ref{fig:analytics}. As can be seen by
comparision with the right subpanel of the left panel of
Fig.~\ref{fig:sRP0}, it agrees well with with our simulation data.

Consistent with this analytical prediction, our numerical results
shown by the colourscale in the right panel of Fig.~\ref{fig:sRP0}
indeed display significant necking post-ramp for values of
$\ebar_0,\edotbar$ above the dot-dashed line. The middle panel of
Fig.~\ref{fig:sRP0} shows the total necking predicted to accumulate
during the straining process itself. As can be seen, ramps with
$\ebar_0,\edotbar$ values that lie above the dot-dashed line but that
also satisfy $\edotbar< 1/\taus$ (effectively recovering
non-stretching Rolie-Poly dynamics) suffer sufficiently strong necking
during the straining process itself that the filament is likely to
fail altogether during straining, rendering any prediction of further
necking post-strain largely irrelevant.  The important new regime lies
at high strain rates $\edotbar \gtrsim 1/\taus$ for imposed strains
$\ebar_0>0.703$, where we find stability against necking during the
straining process itself due to the accumulation of chain stretch, as
discussed earlier in Ref.~\cite{Hoyle2016a}, but with a strong delayed
necking after the straining stops. Ramps with imposed strains
$\ebar_0<0.703$ give no necking, even post-strain.

\section{Nonlinear dynamics}
\label{sec:nonlinear}

\begin{figure*}
\includegraphics[width=0.99\textwidth]{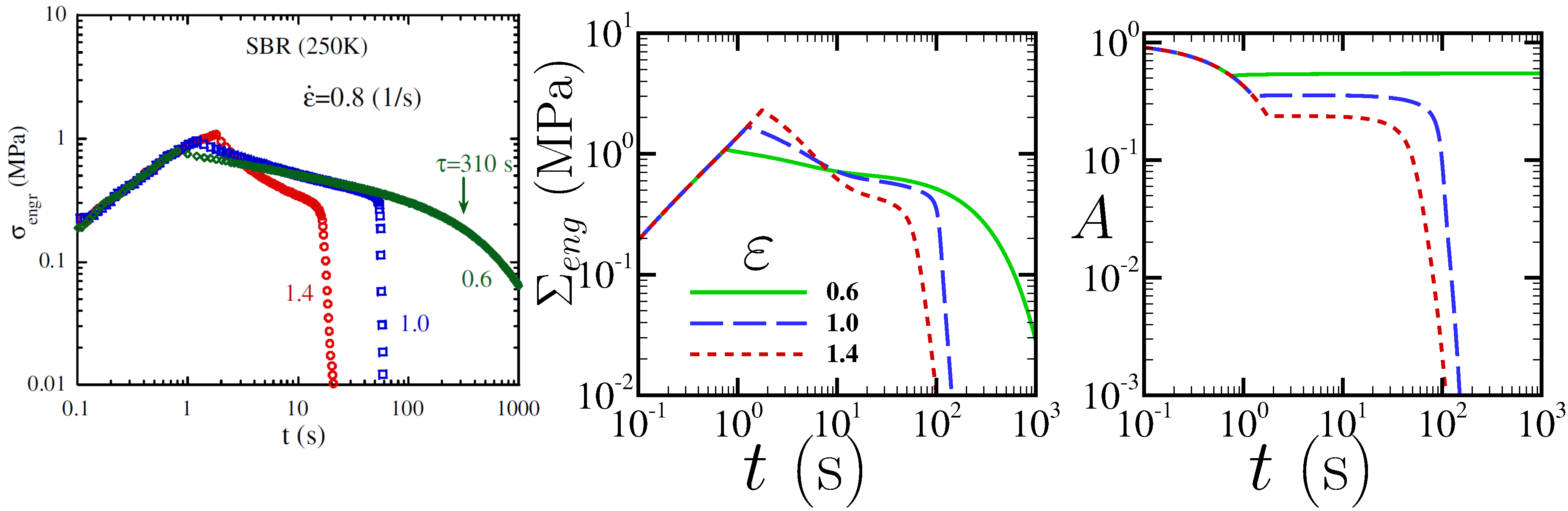}	
\caption{{\bf Left:} Experimental data taken from Ref.~\cite{Wang2007}
  for a monodisperse styrene-butadiene rubber (SBR 250K) for which the
  modulus $G=0.68$MPa, the reptation time $\taud=310$s and the chain
  stretch relaxation time $\taus=4.1$s$=0.0132\taud$. Shown is the
  evolution of the tensile stress during and after three different
  strain ramps: each with an imposed strain rate
  $\edotbar=0.8$s$^{-1}=248/\taud=3.28/\taus$, and with imposed
  strains $\ebar_0=0.6, 1.0$ and $1.4$ respectively. {\bf Middle:}
  Direct counterpart evolution of the stress during strain ramp in our
  nonlinear simulations of the Rolie-Poly model with chain stretch
  included.  Flow protocol and polymeric material parameter values are
  matched to those of the experimental data, and we further assumed a
  value for the Newtonian viscosity $3\eta=0.001G \taud$. {\bf Right:}
  Corresponding simulation data for the area at the mid-point of the
  filament.
    }
\label{fig:nonlinear1} 
\end{figure*}

\begin{figure*}
\includegraphics[width=0.99\textwidth]{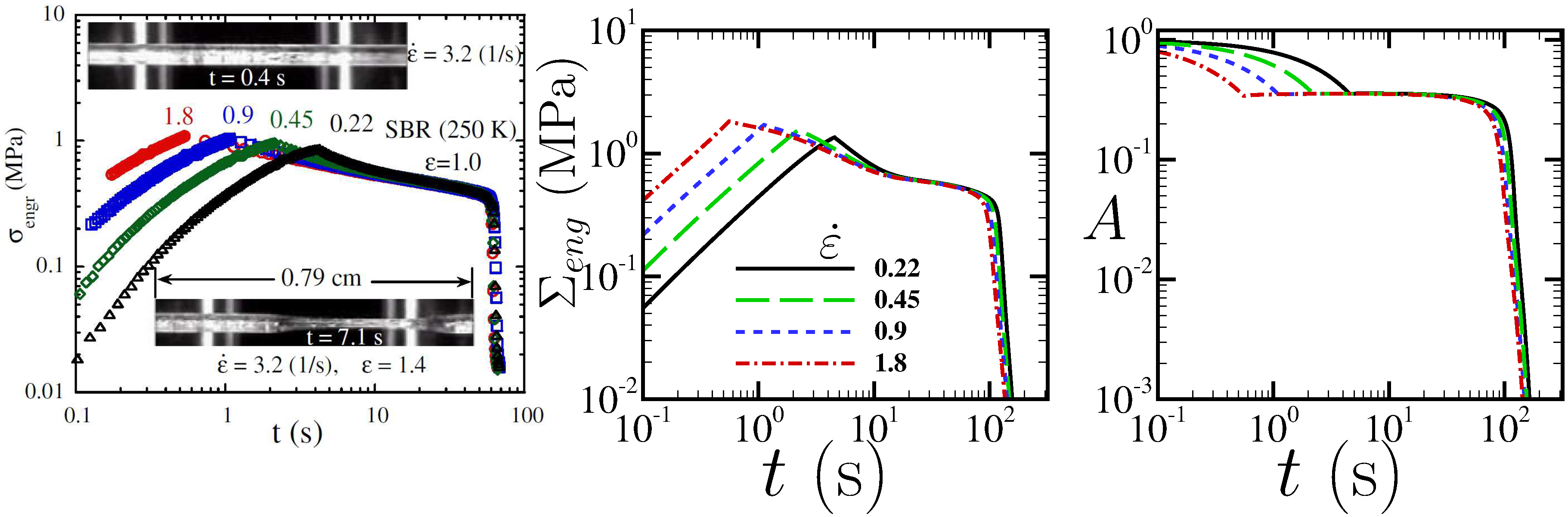}
\caption{{\bf Left:} Experimental data taken from Ref.~\cite{Wang2007}
  for a monodisperse styrene-butadiene rubber (SBR 250K) for which the
  modulus $G=0.68$MPa, the reptation time $\taud=310$s and the chain
  stretch relaxation time $\taus=4.1$s$=0.0132\taud$. Shown is the
  evolution of the tensile stress during and after three different
  strain ramps: each with an imposed strain $\ebar_0=1.0$, and with
  imposed strain rates $\edotbar_0=0.22s^{-1}=68.2/\taud=0.902/\taus$,
  $0.45s^{-1}=140/\taud=1.85/\taus$, $0.9s^{-1}=248/\taud=3.49/\taus$ and
  $1.8s^{-1}=558/\taud=7.38/\taus$ respectively. {\bf Middle:} Direct
  counterpart evolution of the stress during strain ramp in our
  nonlinear simulations of the Rolie-Poly model with chain stretch
  included.  Flow protocol and polymeric material parameter values are
  matched to those of the experimental data, and we further assumed a
  value for the Newtonian viscosity $3\eta=0.001G \taud$.  {\bf Right:}
  Corresponding simulation data for the area at the mid-point of the
  filament.
    }
\label{fig:nonlinear2}		
\end{figure*}

So far, we have discussed the dynamics of necking perturbations to an
initially homogeneous underlying base state in the linear regime, where
the amplitude of the perturbations remains small. We now consider the
dynamics of necking out of that linear regime, when the amplitude of
the perturbations is no longer small. To do so, we numerically evolve
the nonlinear slender filament equations by discretizing them on a
mesh, and time-stepping them using an explicit Euler algorithm for the
spatially local terms and first order upwinding for the convective
terms~\cite{Press2007}.  Details are given in Ref.~\cite{Hoyle2016a},
together with a discussion of convergence on the space and time-steps.

The slender filament approach that we use throughout this work is not
capable of properly implementing the no-slip condition that pertains
where the fluid makes contact with the end-plates. To circumvent this,
we use an approximate mimic of that condition by adopting an
artificially divergent viscosity near each plate, according to
Eqn.~VII.1 of Ref.~\cite{Hoyle2016a}.  This strongly limits the
extensional stretching that can occur in the filament in the vicinity
of each plate, and thereby constrains the filament area to remain close
its initial value at each plate, even as the sample as a whole is
stretched out. This has the effect of forcing the sample to thin more
quickly in the middle than at its ends. This purely geometrical effect
seeds a single neck mid-filament, which is then picked up by the
mechanical necking instability that is our focus.

We perform these simulations in the Rolie-Poly model with chain
stretch. Using this model, we attempt a quantitative comparison with
the experimental data of Ref.~\cite{Wang2007}, focusing in particular
on their monodisperse styrene-butadiene rubber (SBR 250K), which has
modulus $G=0.68$MPa, reptation time $\taud=310$s, and chain stretch
relaxation time $\taus=4.1$s$=0.0132\taud$. For all these non-linear 
simulations we set $3\eta = 0.001G \tau_d$, but also checked that our results 
are robust to reasonable variations in this value. 
In using the Rolie-Poly model we are adopting a course-grained
approximation of an entangled linear polymer chain, whereas a
multimode approach would be needed to capture the full quantitative
behaviour of a real polymeric fluid~\cite{Likhtman2003}.  We therefore
only expect to obtain qualitative agreement with experimental data.

The results are shown in Figs.~\ref{fig:nonlinear1}
and~\ref{fig:nonlinear2}. In Fig.~\ref{fig:nonlinear1} we compare the
results of our theoretical calculations with the experiments of
Ref.~\cite{Wang2007} for the time-evolution of the tensile engineering
stress during and after three different strain ramps, each performed
at an imposed strain rate $\edotbar=0.8$s$^{-1}=248/\taud=3.28/\taus$,
and with imposed strains $\ebar_0=0.6, 1.0$ and $1.4$ respectively.
The middle panel shows the counterpart results for the tensile
engineering stress in our numerical simulations of the Rolie-Poly
model, with model and flow-protocol parameter values matched to those
of the experimental data (and with an assumed solvent viscosity much
smaller than the viscoelastic one, which, as noted above, does not
affect the results). The right panel shows the evolution of the area
at the filament's midpoint, again obtained in our numerical
simulations.  Fig.~\ref{fig:nonlinear2} shows directly corresponding
panels to those in Fig.~\ref{fig:nonlinear1}, but now for four
different strain ramps, each performed to a total strain
$\ebar_0=1.0$, and with imposed strain rates
$\edotbar=0.22s^{-1}=68.2/\taud=0.902/\taus$,
$0.45s^{-1}=140/\taud=1.85/\taus$, $0.9s^{-1}=248/\taud=3.49/\taus$ and
$1.8s^{-1}=558/\taud=7.38/\taus$.  In each case, our numerical simulations
are seen to capture the overall qualitative features of the
experimental data. Full quantitative agreement would not be expected,
due to our use of a simplified constitutive model. 

Each of these ramps shown in Figs.~\ref{fig:nonlinear1}
and~\ref{fig:nonlinear2} lies in the regime where the imposed strain
rate exceeds (or nearly exceeds, for the slowest ramp) the inverse
chain stretch relaxation time. According to our arguments in
Sec.~\ref{sec:linear} above, then, we should expect the filament to be
stabilised against necking during the straining process itself, but
then to display delayed failure during the subsequent process of
stress relaxation, provided the applied strain $\ebar_0>0.703$.
Precisely this scenario is indeed observed: delayed failure is seen in
all the ramps that have $\ebar>0.703$. In contrast, the ramp with
$\ebar_0=0.6$ in Fig.~\ref{fig:nonlinear1} displays full stress
relaxation without filament failure.

\section{Analogy with shear banding}
\label{sec:banding}

In Ref.~\cite{Moorcroft2013}, one of the present authors together with
Moorcroft studied the direct counterpart in shear of the protocol
considered in this manuscript in extension: the imposition of a given
total shear strain $\gamma_0$ at an applied shear strain rate
$\gdotbar$, after which the straining is switched off and the sample
held in its strained state with shear strain $\gamma_0$ for all times
thereafter.  The interest in that work lay in the possibility that
delayed shear banding might arise during the process of stress
relaxation after the straining stops, by analogy with the necking
after an interrupted extensional strain as considered here.  That
delayed shear banding effect had also been considered in the earlier
work of Ref.~\cite{Adams2009}.

Fig.~9 of Ref.~\cite{Moorcroft2013} showed their results for the
stretching Rolie-Poly model in that protocol, for two different values
of the imposed strain rate: one in the regime $1/\taud\ll \gdotbar \ll
1/\taus$ and another in the regime $\gdotbar\gg 1/\taus$. As can be
seen by direct comparison with the left panel of Fig.~\ref{fig:sRP0}
in this work, an analogous scenario with the necking dynamics reported
in this manuscript is indeed evident.  In particular, for the case
$1/\taud\ll \gdotbar \ll 1/\taus$, significant instability to shear
banding was seen both during the straining process itself, and during
the stress relaxation that follows it. In the second case $\gdotbar\gg
1/\taus$, instability to shear banding was suppressed during the
straining process itself.  However significant instability to shear
banding then arose during the first part of the stress decay, on a
timescale $\taus$, associated with the relaxation of chain stretch.
Stability against banding was later recovered on the the longer
timescale $\taud$ on which the chain orientation progressively
relaxed.

\section{Conclusions}
\label{sec:conclusion}

In this work, we have performed linear stability analysis and nonlinear slender filament simulations of extensional necking in complex fluids and soft solids, for the flow protocol in which an initially cylindrical filament (or planar sheet) is subject to a constant extensional strain rate for a given time interval, after which the strain rate is then set to zero. Our focus has been on the conditions required for a sample to neck during the process of stress relaxation after the end of the strain ramp.

We derived analytical criteria for necking during the stress relaxation, within a highly simplified and generalised scalar constitutive model. Within this model we found two different possible modes of necking. The first is associated with an upward curvature in the stress relaxation function shown on a log-linear plot. The second is related to a particular, carefully defined `elastic' derivative of the tensile force with respect to an imagined sudden strain increment.
We showed these two criteria to be in excellent agreement with the behaviour of the Oldroyd B and Giesekus models, and the Rolie-Poly model with chain stretch ignored.

With chain stretch included in the Rolie-Poly model, we obtained a slightly more complicated analytical criterion for necking during the stress relaxation, although with key ingredients closely mirroring counterpart ingredients of the simpler criteria derived within the simplified scalar model.  We showed this criterion to be in full agreement with nonlinear slender filament simulations of the Rolie-Poly model with chain stretch, and with the scenario discussed by the Copenhagen group in Refs.~\cite{Lyhne2009,Rasmussen2011}. In particular, we found delayed necking after ramps with a total accumulated strain exceeding $\ebar\approx 0.7$, for ramp rates exceeding the inverse strain relaxation timescale.

We discussed finally a close analogy between this delayed necking following an interrupted extensional strain ramp and that of delayed shear banding following an interrupted shear strain ramp, as discussed earlier in Ref.~\cite{Moorcroft2013,Adams2009}.

{\it Acknowledgements} -- The research leading to these results has
received funding from the European Research Council under the European
Union's Seventh Framework Programme (FP7/2007-2013) / ERC grant
agreement number 279365.

\section*{Appendix I: correspondence of tensorial and scalar
  constitutive models}

\begin{figure*}
\includegraphics[width=0.99\textwidth]{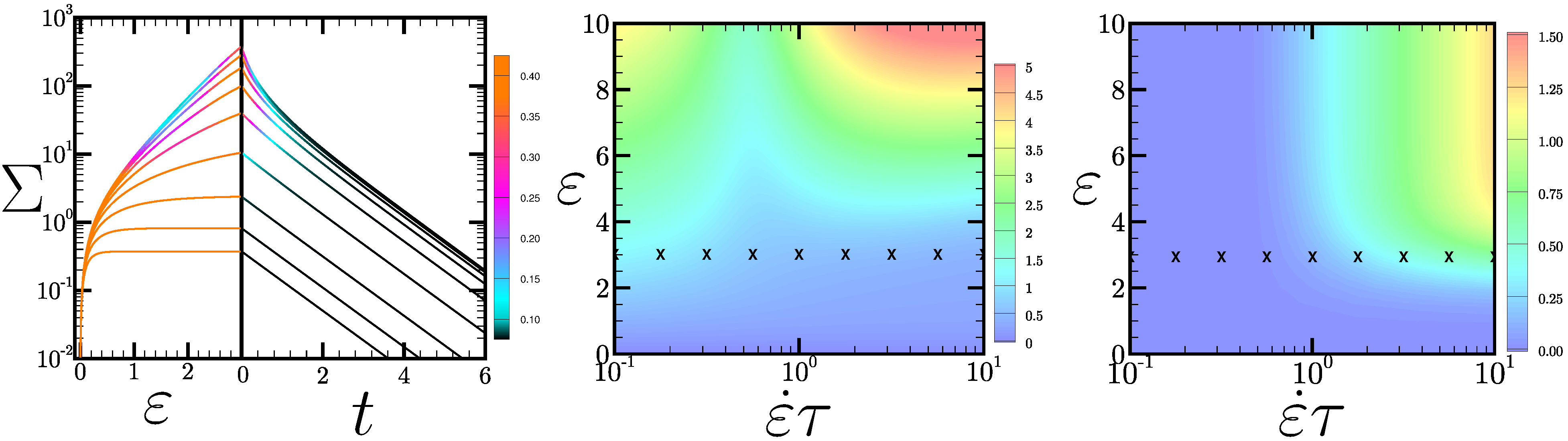}
\caption{Direct counterpart to the numerical results shown for the
  linearised necking dynamics in the fully tensorial Giesekus model
  Fig.~\ref{fig:Giesekus}, now in the scalar equivalent of that
  model.}
  \label{fig:toy} 
\end{figure*}

Our analytical calculations in Sec.~\ref{sec:Giesekus} were performed
in the simplified scalar version of the Giesekus model. In contrast,
our numerical results in Fig.~\ref{fig:Giesekus} are for the fully
tensorial form of the model. To demonstrate the equivalence of these,
in Fig.~\ref{fig:toy} we show numerical results for the scalar version
of the Giesekus model.  Reassuringly, we see close agreement with
those of Fig.~\ref{fig:Giesekus} for the fully tensorial form.

\section*{Appendix II: effect of convective constraint release}

\begin{figure*}
\includegraphics[width=1.0\textwidth]{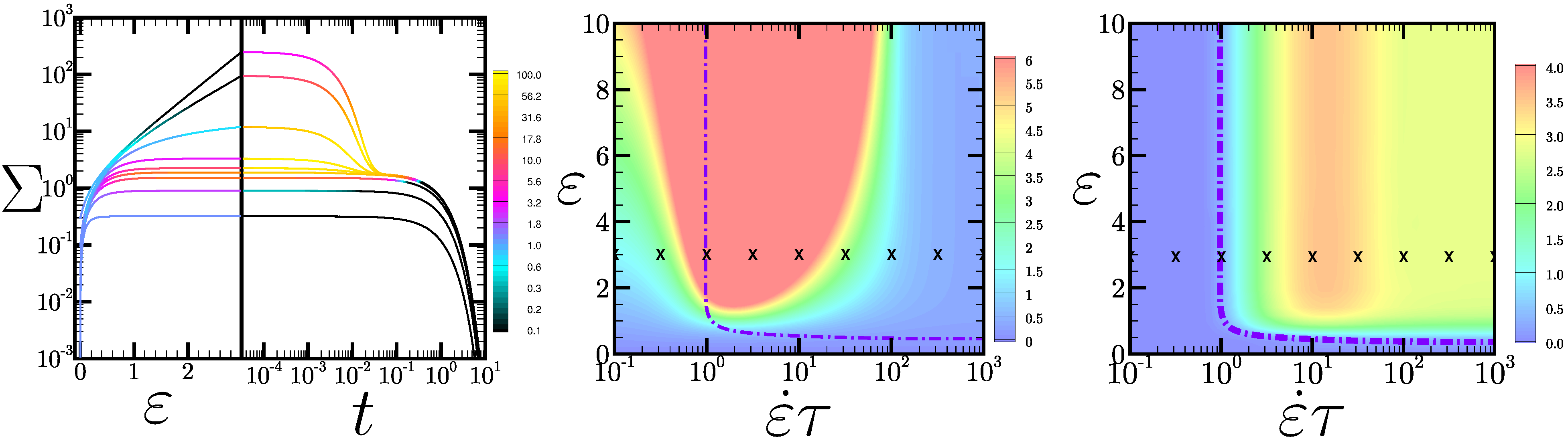}
\caption{Direct counterpart of Fig.~\ref{fig:sRP0} in the main text,
  now for a value of the convective constraint release parameter
  $\beta=1.0$.}
\label{fig:sRP1} 		
\end{figure*}

In Fig.~\ref{fig:sRP0} of the main text, we demonstrated the necking
dynamics post-straining of the Rolie-Poly model with chain stretch
including, for a value of the convective constraint release
parameter $\beta=0.0$. In Fig.~\ref{fig:sRP1}, we show the exactly
counterpart results for a value $\beta=1.0$ at the other end of the
allowed range $0<\beta<1$, demonstrating qualitatively the same
scenario even with convective constraint release included to its
maximum allowable extent.

\end{document}